\documentclass[aps,prd,twocolumn,superscriptaddress,preprintnumbers]{revtex4}
\usepackage{graphicx}

\def \s{\sqrt{2}}
\def \st{\sqrt{3}}
\def \sx{\sqrt{6}}

\def\be{\begin{equation}}
\def\ee{\end{equation}}
\def\bea{\begin{eqnarray}}
\def\eea{\end{eqnarray}}
\def\bean{\begin{eqnarray*}}
\def\eean{\end{eqnarray*}}
\def\bary{\begin{array}}
\def\eary{\end{array}}

\def\bit{\begin{itemize}}
\def\eit{\end{itemize}}
\def\bwt{\begin{widetext}}
\def\ewt{\end{widetext}}
\def\half{\frac{1}{2}}
\def \cth{c_\theta}
\def \sth{s_\theta}

\def\ol{\overline}

\def\cB{{\cal{B}}}

\def\vud{V_{ud}}
\def\vus{V_{us}}
\def\vub{V_{ub}}

\def\vtd{V_{td}}
\def\vts{V_{ts}}
\def\vtb{V_{tb}}
\def\ub{{\bar u}}
\def\cb{{\bar c}}

\def\db{{\bar d}}
\def\sb{{\bar s}}
\def\bb{{\bar b}}
\begin{document}

\preprint{\vbox{ \hbox{ANL-HEP-PR-03-037}
                 \hbox{EFI-03-24}
                 \hbox{hep-ph/0306021}
                 \hbox{June 2003}
         }}

\title{TWO-BODY CHARMLESS $B$ DECAYS INVOLVING $\eta$ AND $\eta'$
\footnote{To be submitted to Phys.\ Rev.\ D.}}

\author{Cheng-Wei Chiang}
\email[e-mail: ]{chengwei@hep.uchicago.edu}
\affiliation{Enrico Fermi Institute and Department of Physics,
University of Chicago, 5640 S. Ellis Avenue, Chicago, IL 60637}
\affiliation{HEP Division, Argonne National Laboratory,
9700 S. Cass Avenue, Argonne, IL 60439}
\author{Michael Gronau}
\email[e-mail: ]{gronau@physics.technion.ac.il}
\affiliation{Enrico Fermi Institute and Department of Physics, 
University of Chicago, 5640 S. Ellis Avenue, Chicago, IL 60637}
\affiliation{Department of Physics, Technion -- Israel Institute of
Technology, Haifa 32000, Israel (Permanent Address)}
\author{Jonathan L.~Rosner}
\email[e-mail: ]{rosner@hep.uchicago.edu}
\affiliation{Enrico Fermi Institute and Department of Physics, 
University of Chicago, 5640 S. Ellis Avenue, Chicago, IL 60637}

\date{\today}

\begin{abstract}
  We discuss implications of recent experimental data for $B$ decays into two
  pseudoscalar mesons, with emphasis on those with $\eta$ and $\eta'$ in the
  final states.  Applying a U-spin argument, we show that tree and penguin
  amplitudes, both in $B^+ \to \pi^+ \eta$ and in $B^+ \to \pi^+ \eta'$, are of
  comparable magnitudes.  Nontrivial relative weak and strong phases between
  the tree-level amplitudes and penguin-loop amplitudes in the $B^{\pm} \to
  \pi^{\pm} \eta$ modes are extracted.  We predict possible values for the
  averaged branching ratio and $CP$ asymmetry of the $B^{\pm} \to \pi^{\pm}
  \eta'$ modes.
  We test the assumption of a singlet-penguin amplitude with the same weak and
  strong phases as the QCD penguin in explaining the large branching ratios of
  $\eta' K$ modes, and show that it is consistent with current branching ratio
  and $CP$ asymmetry data of the $B^+ \to (\pi^0, \eta, \eta') K^+$ modes.  We
  also show that the strong phases of the singlet-penguin and tree-level
  amplitudes can be extracted with further input of electroweak penguin
  contributions and a sufficiently well-known branching ratio of the $\eta K^+$
  mode.  Using $SU(3)$ flavor symmetry, we also estimate required data
  samples to detect modes that have not yet been seen.
\end{abstract}

\pacs{13.25.Hw, 14.40.Nd, 11.30.Er, 11.30.Hv}

\maketitle

\section{INTRODUCTION \label{sec:intro}}

The KEK-B and PEP-II $e^+ e^-$ colliders and Belle and BaBar detectors have
permitted the study of $B$ decays with unprecedented sensitivity.
$CP$-violating asymmetries in $B^0 \to J/\psi K_S$ and related modes have been
observed \cite{Babeta,Bebeta} and agree with predictions based on the
Kobayashi-Maskawa theory \cite{KM}.  These asymmetries are associated with the
interference between $B^0$--$\bar B^0$ mixing and a single decay amplitude.
The observation of {\it direct} $CP$ asymmetries in $B$ decays, associated with
two amplitudes differing in both weak and strong phases, has remained elusive.
In this article we demonstrate that the data on $B \to PP$ branching ratios,
where $P$ denotes a pseudoscalar meson, now indicate that substantial direct
$CP$ asymmetries in the decays $B^+ \to \pi^+ \eta$ and $B^+ \to \pi^+ \eta'$,
anticipated previously \cite{BRS,DGR,AK}, are likely.  Indeed, a recent BaBar
result \cite{Aubert:2003ez} favors a large $\pi^+ \eta$ asymmetry.

We shall discuss $B^0 \to PP$ and $B^+ \to PP$ decays within the framework of
$SU(3)$ flavor symmetry \cite{DZ,SW,Chau,Gronau:1994rj,GR95,Grinstein:1996us},
introducing corrections for $SU(3)$
breaking or assigning appropriate uncertainties.  Our treatment will be an
update of previous discussions \cite{VPUP,Chiang:2001ir}, to which we refer for
further details.  We shall be concerned here mainly with the decays of charged
and neutral $B$ mesons to $K \eta$, $K \eta'$, $\pi \eta$, and $\pi \eta'$.  We
shall compare our results with a recent treatment also based on flavor $SU(3)$
symmetry \cite{FHH}.

In Section \ref{sec:not} we review notation and amplitude decompositions using
flavor symmetry.  We compare these with experimental rates, obtaining
magnitudes of amplitudes, in Section \ref{sec:amp}.  We can then extract
amplitudes corresponding to specific flavor topologies in Section
\ref{sec:extract}.  Section \ref{sec:Keta} is devoted to $B \to \eta K$ and $B
\to \eta' K$, while discussions of $B^+ \to \pi^+ \eta$ and $B^+ \to \pi^+
\eta'$ occupy Section \ref{sec:pi-eta}.  Some progress on testing amplitude
relations proposed in Refs.\ \cite{GR95,DGR} is noted in Section
\ref{sec:amprel}.  Relations among all charged $B$ decays, obtained by applying
only the U-spin subgroup \cite{MLL,Uspin} of flavor $SU(3)$, are studied in
Section \ref{sec:uspin}.  We remark on as yet unseen processes such as $B^+ \to
K^+ \ol K^0$ and $B^0 \to (K^0 \ol K^0, ~\pi^0\pi^0,~\pi^0 \eta,~\pi^0 \eta')$
in Section \ref{sec:unseen}, and conclude in Section \ref{sec:summary}.  An
Appendix compares our methods with those used in Ref.\ \cite{GLNQ} to estimate
non-penguin contributions to $B^0 \to \eta' K^0$.

\section{NOTATION \label{sec:not}}

Our quark content and phase conventions \cite{Gronau:1994rj,GR95} are:
\begin{itemize}
\item{ {\it Bottom mesons}: $B^0=d\bb$, ${\ol B}^0=b\db$, $B^+=u\bb$,
    $B^-=-b\ub$, $B_s=s\bb$, ${\ol B}_s=b\sb$;}
\item{ {\it Charmed mesons}: $D^0=-c\ub$, ${\ol D}^0=u\cb$, $D^+=c\db$,
    $D^-=d\cb$, $D_s^+=c\sb$, $D_s^-=s\cb$;}
\item{ {\it Pseudoscalar mesons}: $\pi^+=u\db$, $\pi^0=(d\db-u\ub)/\sqrt{2}$,
    $\pi^-=-d\ub$, $K^+=u\sb$, $K^0=d\sb$, ${\ol K}^0=s\db$, $K^-=-s\ub$,
    $\eta=(s\sb-u\ub-d\db)/\sqrt{3}$,
    $\eta^{\prime}=(u\ub+d\db+2s\sb)/\sqrt{6}$;}
\end{itemize}
The $\eta$ and $\eta'$ correspond to octet-singlet mixtures
\be
\eta  =   \eta_8 \cos \theta_0 + \eta_1 \sin \theta_0~,~~
\eta' = - \eta_8 \sin \theta_0 + \eta_1 \cos \theta_0~~,
\ee
with $\theta_0 = \sin^{-1}(1/3) = 19.5^\circ$.

In the present approximation there are seven types of independent amplitudes: a
``tree'' contribution $t$; a ``color-suppressed'' contribution $c$; a
``penguin'' contribution $p$; a ``singlet penguin'' contribution $s$, in which
a color-singlet $q \bar q$ pair produced by two or more gluons or by a $Z$ or
$\gamma$ forms an $SU(3)$ singlet state; an ``exchange'' contribution $e$, an
``annihilation'' contribution $a$, and a ``penguin annihilation'' contribution
$pa$.  These amplitudes contain both the leading-order and electroweak penguin
contributions:
\be\bary{lll}
\label{eqn:dict}
t \equiv T + P_{\rm EW}^C ~, &\quad& c \equiv C + P_{\rm EW} ~, \\
p \equiv P - \frac{1}{3} P_{\rm EW}^C ~, &\quad&
s \equiv S - \frac{1}{3} P_{\rm EW} ~, \\
a \equiv A ~, &\quad& e + pa \equiv E + PA ~,
\eary\ee
where the capital letters denote the leading-order contributions
\cite{Gronau:1994rj,GR95,DGR,Gronau:1995hn} while $P_{\rm EW}$ and
$P_{\rm EW}^C$ are respectively color-favored and color-suppressed electroweak
penguin amplitudes \cite{Gronau:1995hn}.  We shall neglect smaller terms
\cite{EWVP,GR2001} $PE_{\rm EW}$ and $PA_{\rm EW}$ [$(\gamma,Z)$-exchange and
$(\gamma,Z)$-direct-channel electroweak penguin amplitudes].  We shall denote
$\Delta S = 0$ transitions by unprimed quantities and $|\Delta S| = 1$
transitions by primed quantities.  The hierarchy of these amplitudes can be
found in Ref.~\cite{Chiang:2001ir}.

The partial decay width of two-body $B$ decays is
\be
\label{eq:width}
\Gamma(B \to M_1 M_2)
= \frac{p_c}{8 \pi m_B^2} |{\cal A}(B \to M_1 M_2)|^2 ~,
\ee
where $p_c$ is the momentum of the final state meson in the rest frame of $B$,
$m_B$ is the $B$ meson mass, and $M_1$ and $M_2$ can be either pseudoscalar or
vector mesons.  Using Eq.~(\ref{eq:width}), one can extract the invariant
amplitude of each decay mode from its experimentally measured branching ratio.
To relate partial widths to branching ratios, we use the world-average
lifetimes $\tau^+ = (1.656\pm0.014)$ ps and $\tau^0 = (1.539\pm0.014)$ ps
computed by the LEPBOSC group \cite{LEPBOSC}.  Unless otherwise indicated, for
each branching ratio quoted we imply the average of a process and its
$CP$-conjugate.

\section{AMPLITUDE DECOMPOSITIONS AND EXPERIMENTAL RATES \label{sec:amp}}

The experimental branching ratios and $CP$ asymmetries on which our analysis is
based are listed in Tables \ref{tab:dS0data} and \ref{tab:dS1data}.
Contributions from the CLEO \cite{Richichi:1999kj,Bornheim:2003bv}, BaBar
\cite{Aubert:2003ez,Aubert:2003qj,Aubert:2002ng,Aubert:2001zf,Aubert:2002jb,%
  Aubert:2001hs,Aubert:2003bq,Aubert:2002jm}, and Belle
\cite{Tomura,Abe:2001pf,Abe:2003ja,Unno:2003,Chen:2002af,Abe:2002np}
Collaborations are included.  In addition we shall make use of the 90\% c.l.\ 
upper bounds \cite{Hagiwara:fs} $\ol{\cal B}(B^0 \to \eta \eta,~\eta
\eta',~\eta' \eta') < (18,~27,~47) \times 10^{-6}$.

\begin{table*}
\caption{Experimental branching ratios of selected $\Delta S = 0$ decays of $B$
  mesons.  Branching ratios are quoted in units of $10^{-6}$.  Numbers in
  parentheses are upper bounds at 90 \% c.l.  References are given in square
  brackets.  Additional lines, if any, give the $CP$ asymmetry ${\cal A}_{CP}$
  (second line) or $({\cal S},{\cal A})$ (second and third lines) for charged
  or neutral modes, respectively.
\label{tab:dS0data}}
\begin{ruledtabular}
\begin{tabular}{llllll}
 & Mode & CLEO & BaBar & Belle & Average \\ 
\hline
$B^+ \to$
    & $\pi^+ \pi^0$ 
        & $4.6^{+1.8+0.6}_{-1.6-0.7}$ \cite{Bornheim:2003bv}
        & $5.5^{+1.0}_{-0.9}\pm0.6$ \cite{Aubert:2003qj}
        & $5.3\pm1.3\pm0.5$ \cite{Tomura}
        & $5.27\pm0.79$ \\
    &   & -
        & $-0.03^{+0.18}_{-0.17}\pm0.02$ \cite{Aubert:2003qj}
        & $-0.14\pm0.24^{+0.05}_{-0.04}$ \cite{Tomura}
        & $-0.07\pm0.14$ \\
    & $K^+ \ol{K}^0$ 
        & $<3.3$ \cite{Bornheim:2003bv}
        & $-0.6^{+0.6}_{-0.7}\pm0.3 \; (<1.3)$ \cite{Aubert:2002ng}
        & $1.7\pm1.2\pm0.1 \; (<3.4)$ \cite{Tomura}
        & $<1.3$ \\
    & $\pi^+ \eta$ 
        & $1.2^{+2.8}_{-1.2} \; (<5.7)$ \cite{Richichi:1999kj}
        & $4.2^{+1.0}_{-0.9}\pm0.3$ \cite{Aubert:2003ez}
        & $5.2^{+2.0}_{-1.7} \pm 0.6$ \cite{Tomura}
        & $4.12\pm0.85$ \\
    &   & - 
        & $-0.51^{+0.20}_{-0.18}\pm0.01$  \cite{Aubert:2003ez}
        & -
        & $-0.51\pm0.19$ \\
    & $\pi^+ \eta'$ 
        & $1.0^{+5.8}_{-1.0} \; (<12)$ \cite{Richichi:1999kj}
        & $5.4^{+3.5}_{-2.6}\pm0.8 \; (<12)$ \cite{Aubert:2001zf}
        & $<7$ \cite{Abe:2001pf}
        & $<7$ \\
\hline
$B^0 \to$
    & $\pi^+ \pi^-$ 
        & $4.5^{+1.4+0.5}_{-1.2-0.4}$ \cite{Bornheim:2003bv}
        & $4.7\pm0.6\pm0.2$ \cite{Aubert:2002jb}
        & $4.4\pm0.6\pm0.3$ \cite{Tomura}
        & $4.55\pm0.44$ \\
    &   & -
        & $(0.02\pm0.34\pm0.05,$ \cite{Aubert:2002jb}
        & $(-1.23\pm0.41^{+0.08}_{-0.07},$ \cite{Abe:2003ja}
        & $(-0.49\pm0.27,$ \\
    &   & -
        & $0.30\pm0.25\pm0.04)$ \cite{Aubert:2002jb}
        & $0.77\pm0.27\pm0.08)$ \cite{Abe:2003ja}
        & $0.51\pm0.19)$ \\
    & $\pi^0 \pi^0$
        & $<4.4$ \cite{Bornheim:2003bv}
        & $1.6^{+0.7+0.6}_{-0.6-0.3} \; (<3.6)$ \cite{Aubert:2003qj} 
        & $1.8^{+1.4+0.5}_{-1.3-0.7} \; (<4.4)$ \cite{Tomura}
        & $<3.6$ \\
    & $K^+ K^-$ 
        & $<0.8$ \cite{Bornheim:2003bv}
        & $<0.6$ \cite{Aubert:2002jb}
        & $<0.7$ \cite{Tomura}
        & $<0.6$ \\
    & $K^0 \ol{K}^0$ 
        & $<3.3$ \cite{Bornheim:2003bv}
        & $<2.4$ \cite{Aubert:2001hs} 
        & $0.8\pm0.8\pm0.1 \; (<3.2)$ \cite{Tomura}
        & $<2.4$ \\
    & $\pi^0 \eta$ 
        & $0.0^{+0.8}_{-0.0} \; (<2.9)$ \cite{Richichi:1999kj}
        & - & -
        & $<2.9$ \\
    & $\pi^0 \eta'$ 
        & $0.0^{+1.8}_{-0.0} \; (<5.7)$ \cite{Richichi:1999kj} 
        & - & -
        & $<5.7$ \\
\end{tabular}
\end{ruledtabular}
\end{table*}
%

\begin{table*}
\caption{Same as Table \ref{tab:dS1data} for $|\Delta S| = 1$ decays of $B$
 mesons.
\label{tab:dS1data}}
\begin{ruledtabular}
\begin{tabular}{llllll}
 & Mode & CLEO & BaBar & Belle & Average \\ 
\hline
$B^+ \to$
    & $\pi^+ K^0$ 
        & $18.8^{+3.7+2.1}_{-3.3-1.8}$ \cite{Bornheim:2003bv}
        & $17.5^{+1.8}_{-1.7}\pm1.3$ \cite{Aubert:2002ng}
        & $22.0\pm1.9\pm1.1$ \cite{Tomura}
        & $19.61\pm1.44$ \\
    &   & -
        & $-0.17\pm0.10\pm0.02$ \cite{Aubert:2002ng}
        & $0.07^{+0.09+0.01}_{-0.08-0.03}$ \cite{Unno:2003}
        & $-0.032\pm0.066$ \\
    & $\pi^0 K^+$ 
        & $12.9^{+2.4+1.2}_{-2.2-1.1}$ \cite{Bornheim:2003bv}
        & $12.8^{+1.2}_{-1.1}\pm1.0$ \cite{Aubert:2003qj}
        & $12.8\pm1.4^{+1.4}_{-1.0}$ \cite{Tomura}
        & $12.82\pm1.07$ \\
    &   & -
        & $-0.09\pm0.09\pm0.01$ \cite{Aubert:2003qj}
        & $0.23\pm0.11^{+0.01}_{-0.04}$ \cite{Tomura}
        & $0.035\pm0.071$ \\
    & $\eta K^+$ 
        & $2.2^{+2.8}_{-2.2} \; (<6.9)$ \cite{Richichi:1999kj}
        & $2.8^{+0.8}_{-0.7}\pm0.2$ \cite{Aubert:2003ez}
        & $5.3^{+1.8}_{-1.5}\pm0.6$ \cite{Tomura} 
        & $3.15\pm0.69$ \\
    &   & -
        & $-0.32^{+0.22}_{-0.18}\pm0.01$ \cite{Aubert:2003ez}
        & -
        & $-0.32\pm0.20$ \\
    & $\eta' K^+$ 
        & $80^{+10}_{-9}\pm7$ \cite{Richichi:1999kj}
        & $76.9\pm3.5\pm4.4$ \cite{Aubert:2003bq}
        & $78\pm6\pm9$ \cite{Tomura}
        & $77.57\pm4.59$ \\
    &   & -
        & $0.037\pm0.045\pm0.011$ \cite{Aubert:2003bq}
        & $-0.015\pm0.070\pm0.009$ \cite{Chen:2002af}
        & $-0.002\pm0.040$ \\
\hline
$B^0 \to$
    & $\pi^- K^+$ 
        & $18.0^{+2.3+1.2}_{-2.1-0.9}$ \cite{Bornheim:2003bv}
        & $17.9\pm0.9\pm0.7$ \cite{Aubert:2002jb}
        & $18.5\pm1.0\pm0.7$ \cite{Tomura}
        & $18.16\pm0.79$ \\
    &   & -
        & $-0.102\pm0.050\pm0.016$ \cite{Aubert:2002jb}
        & $-0.07\pm0.06\pm0.01$ \cite{Tomura}
        & $-0.088\pm0.040$ \\
    & $\pi^0 K^0$ 
        & $12.8^{+4.0+1.7}_{-3.3-1.4}$ \cite{Bornheim:2003bv}
        & $10.4\pm1.5\pm0.8$ \cite{Aubert:2002jm}
        & $12.6\pm2.4\pm1.4$ \cite{Tomura}
        & $11.21\pm1.36$ \\
    &   & -
        & $0.03\pm0.36\pm0.09$ \cite{Aubert:2002jm}
        & -
        & $0.03\pm0.37$ \\
    & $\eta K^0$ 
        & $0.0^{+3.2}_{-0.0} \; (<9.3)$ \cite{Richichi:1999kj}
        & $2.6^{+0.9}_{-0.8}\pm0.2 \; (<4.6)$ \cite{Aubert:2003ez} 
        & $<12$ \cite{Tomura}
        & $<4.6$ \\
    & $\eta' K^0$ 
        & $89^{+18}_{-16}\pm9$ \cite{Richichi:1999kj}
        & $55.4\pm5.2\pm4.0$ \cite{Aubert:2003bq}
        & $68\pm10^{+9}_{-8}$ \cite{Tomura} 
        & $60.57\pm5.61$ \\
    &   & - 
        & $(0.02\pm0.34\pm0.03,$ \cite{Aubert:2003bq}
        & $(0.71\pm0.37^{+0.05}_{-0.06}$ \cite{Abe:2002np}
        & $(0.33\pm0.25,$ \\
    &   & -
        & $-0.10\pm0.22\pm0.03)$ \cite{Aubert:2003bq}
        & $-0.26\pm0.22\pm0.03$ \cite{Abe:2002np}
        & $-0.18\pm0.16)$ \\
\end{tabular}
\end{ruledtabular}
\end{table*}

We list theoretical predictions and averaged experimental data for charmless $B
\to PP$ decays involving $\Delta S = 0$ transitions in Table \ref{tab:dS0} and
those involving $|\Delta S| = 1$ transitions in Table \ref{tab:dS1}.  Numbers
in italics are assumed inputs. All other numbers are inferred using additional
assumptions and $SU(3)_F$-breaking and CKM factors.  Terms of order $\lambda^2$
and smaller relative to dominant amplitudes are omitted.  These results update
ones quoted most recently in Ref.\ \cite{Chiang:2001ir}.  The magnitudes of
individual amplitudes are based on predicted values (see Table
\ref{tab:amp-value} below) and include the appropriate Clebsch-Gordan
coefficients for each mode.

\begin{table*}
\caption{Summary of predicted contributions to $\Delta S = 0$ decays of $B$
  mesons to two pseudoscalars.  Amplitude magnitudes $|A_{\rm exp}|$ extracted
  from experiments are quoted in units of $10^{-9}$GeV.  Numbers in italics are
  assumed inputs.  Others are inferred using additional assumptions and
  $SU(3)_F$-breaking and CKM factors.
\label{tab:dS0}}
\begin{ruledtabular}
\begin{tabular}{llccccccc}
 & Mode & Amplitudes & $|t+c|$ & $|p|$ & $|s|$ \footnotemark[1]
 & $p_c$ (GeV) & $|A_{\rm exp}|$ & ${\cal A}_{CP}$ \\ 
\hline
$B^+ \to$
    & $\pi^+ \pi^0$ & $-\frac{1}{\sqrt{2}}(t+c)$
        & {\it 23.59} & 0 & 0 & $2.636$ & $23.59\pm1.76$ & $-0.07\pm0.14$ \\
    & $K^+ \ol{K}^0$ & $p$
        & 0 & 9.00 & 0 & $2.593$ & $<11.82$ \\
    & $\pi^+ \eta$ & $-\frac{1}{\sqrt{3}}(t+c+2p+s)$ 
        & 19.26 & 10.39 & 2.15 & $2.609$ & $20.95\pm2.15$ & $-0.51\pm0.19$ \\
    & $\pi^+ \eta'$ & $\frac{1}{\sqrt{6}}(t+c+2p+4s)$ 
        & 13.62 & 7.35 & 6.09 & $2.551$ & $<27.64$ \\
\hline
$B^0 \to$
    & $\pi^+ \pi^-$ & $-(t+p)$
        & {\it 27.12} \footnotemark[2] & 9.01 & 0 & $2.636$ & $22.73\pm1.09$
        & $({\cal S},{\cal A})$ \footnotemark[3] \\
    & $\pi^0 \pi^0$ & $-\frac{1}{\sqrt{2}}(c-p)$ 
        & - & 6.36 & 0 & $2.636$ & $<20.21$ \\
    & $K^+ K^-$ & $-(e+pa)$ 
        & 0 & 0 & 0 & $2.593$ & $<8.32$ \\
    & $K^0\ol{K}^0$ & $p$ 
        & 0 & 9.00 & 0 & $2.592$ & $<16.64$ \\
    & $\pi^0\eta$ & $-\frac{1}{\sqrt{6}}(2p+s)$ 
        & - & 7.35 & 1.52 & $2.610$ & $<18.25$ \\
    & $\pi^0\eta'$ & $\frac{1}{\sqrt{3}}(p+2s)$ 
        & - & 5.20 & 4.30 & $2.551$ & $<25.87$ \\
    & $\eta\eta$ & $\frac{\sqrt{2}}{3}(c+p+s)$ 
        & - & 4.24 & 1.76 & $2.582$ & $<45.70$ \\
    & $\eta\eta'$ & $-\frac{\sqrt{2}}{3}(c+p+\frac52s)$ 
        & - & 4.24 & 4.40 & $2.523$ & $<56.63$ \\
    & $\eta'\eta'$ & $\frac{1}{3\sqrt{2}}(c+p+4s)$ 
        & - & 2.12 & 3.52 & $2.460$ & $<75.66$ \\
\end{tabular}
\end{ruledtabular}
\footnotetext[1]{\leftline{Assuming constructive interference between $s'$ and
  $p'$ in $B \to \eta' K$ (Table \ref{tab:dS1}).}}
\footnotetext[2]{$T \simeq t$ contribution alone.}
\footnotetext[3]{$({\cal S},{\cal A}) = (-0.49\pm0.27,0.51\pm0.19)$.}
\end{table*}
%

\begin{table*}
\caption{Same as Table \ref{tab:dS0} for $|\Delta S| = 1$ decays of $B$ mesons.
\label{tab:dS1}}
\begin{ruledtabular}
\begin{tabular}{llccccccc}
 & Mode & Amplitudes & $|T'+C'|$ & $|p'|$ & $|s'|$ \footnotemark[1]
 & $p_c$ (GeV) & $|A_{\rm exp}|$ & ${\cal A}_{CP}$ \\ 
\hline
$B^+ \to$
    & $\pi^+ K^0$ & $p'$ 
        & 0 & {\it 45.70} & 0 & $2.614$ & $45.70\pm1.68$ & $-0.032\pm0.066$ \\
    & $\pi^0 K^+$ & $-\frac{1}{\sqrt{2}}(p'+t'+c')$ 
        & 6.61 & 32.32 & 0 & $2.615$ & $36.94\pm1.54$ & $0.035\pm0.071$ \\
    & $\eta K^+$ & $-\frac{1}{\sqrt{3}}(s'+t'+c')$
        & 5.40 & 0 & 10.92 & $2.588$ & $18.40\pm2.01$ & $-0.32\pm0.20$ \\
    & $\eta' K^+$ & $\frac{1}{\sqrt{6}}(3p'+4s'+t'+c')$ 
        & 3.82 & 55.97 & 30.88 & $2.530$ & $92.42\pm2.74$ 
        & $-0.002\pm0.040$ \\
\hline
$B^0 \to$
    & $\pi^- K^+$ & $-(p'+t')$
        & 7.59 \footnotemark[2] & 45.70 & 0 & $2.615$ 
        & $45.57\pm0.99$ & $-0.088\pm0.040$ \\
    & $\pi^0 K^0$ & $\frac{1}{\sqrt{2}}(p'-c')$ 
        & - & 32.32 & 0 & $2.614$ & $35.81\pm2.17$ \\
    & $\eta K^0$ & $-\frac{1}{\sqrt{3}}(s'+c')$
        & - & 0 & 10.92 & $2.587$ & $<23.06$ \\
    & $\eta' K^0$ & $\frac{1}{\sqrt{6}}(3p'+4s'+c')$
        & - & 55.97 & 30.88 & $2.528$ & $84.73\pm3.93$
        & $({\cal S},{\cal A})$ \footnotemark[3] \\
\end{tabular}
\end{ruledtabular}
\footnotetext[1]{\leftline{Assuming constructive interference between $s'$ and
    $p'$ in $B \to \eta' K$.}}
\footnotetext[2]{$T'$ contribution alone.}
\footnotetext[3]{$({\cal S},{\cal A}) = (0.02\pm0.34,-0.10\pm0.22)$.}
\end{table*}
%

\section{EXTRACTING AMPLITUDES \label{sec:extract}}

We begin with those amplitudes or combinations for which information is
provided by a single decay or by an independent analysis.  We then indicate how
the remaining amplitudes may be determined or bounded.  We expect $p'$, $t+c$,
and $s'$ to dominate most decays in which they occur, while $p$, $t'+c'$, and
$s$ should be of relative order $\lambda$ with respect to them.

The decay $B^+ \to \pi^+ K^0$ is expected to be dominated by the amplitude
$|p'|$ aside from a very small annihilation contribution, as shown in Table
\ref{tab:dS1}.  We thus extract $|p'| = (45.7 \pm 1.7) \times 10^{-9}$ GeV from
the $B^+ \to \pi^+ K^0$ branching ratio.

In principle, $|p|$ for $\Delta S = 0$ transitions could be directly obtained
from the $B^+ \to K^+ \ol K^0$ and $B^0 \to K^0 \ol K^0$ modes.  However,
current experiments only give upper bounds on their branching ratios.  Instead,
we use the relation $|p/p'| = |\vtd / \vts| = \lambda |1 - \ol \rho - i \ol
\eta|$, assuming both $p$ and $p'$ to be dominated by the top quark loop.  The
central values $(\overline \rho,~\overline \eta) = (0.21,0.34)$ quoted in one
analysis \cite{CKMfitter}, together with their 68\% c.l. limits, imply $|p/p'|
= 0.197 \pm 0.012$ for $\lambda = 0.2240$ (see \cite{Battaglia:2003in}), and
hence $|p| = (9.00 \pm 0.64) \times 10^{-9}$ GeV.  Although this is the nominal
$1 \sigma$ error, the range $\ol \rho \in [0.08,0.34]$, $\ol \eta \in
[0.25,0.43]$ quoted in Ref.\ \cite{CKMfitter} implies an error for $|p|$ more
like 20\% when theoretical uncertainties affecting $\ol \rho$ and $\ol \eta$
are taken into account.  We shall see that the prospects are good for reducing
this error by direct measurement of the $K \ol K$ branching ratios mentioned
above.

In the majority of our discussion we will be using a convention in which
penguin amplitudes are governed by CKM factors $V^*_{tb}V_{ts}$ and
$V^*_{tb}V_{td}$, corresponding to strangeness changing and strangeness
conserving decays, respectively. In an alternative convention \cite{conv} one
integrates out the top quark in the $\bar b \to \bar s (\bar d)$ loops and
uses the unitarity relations $V^*_{tb}V_{ts(d)} = - V^*_{cb}V_{cs(d)} -
V^*_{ub}V_{us(d)}$. In this convention penguin amplitudes are governed by
$V^*_{cb}V_{cs}$ and $V^*_{cb}V_{cd}$. The ratio of these CKM factors is better
known than that occuring in the other convention. However $SU(3)$ breaking
corrections, possibly of the form $f_K/f_\pi$ would introduce an uncertainty of
about 20\% in $|p/p'|$, similar to the above. We will return to this convention
when discussing the consequences of U-spin symmetry in Section \ref{sec:uspin}.

Another combination which can be extracted directly from data is $t+c$.  The
electroweak penguin contribution to this amplitude is expected to be small and
we shall neglect it.  The average branching ratio $\ol{\cal B}(B^+ \to \pi^+
\pi^0) = (5.27 \pm 0.79) \times 10^{-6}$ quoted in Table \ref{tab:dS0data}
gives $|t+c| = (33.3 \pm 2.5) \times 10^{-9}$ GeV. Two subsequent routes permit
the separate determination of $t$ and $c$.

Factorization calculations \cite{BBNS} in principle can yield the ratio $|C/T|$
of leading color-suppressed to color-favored amplitudes [exclusive of the
electroweak penguin amplitudes in Eq.\ (\ref{eqn:dict})].  However, at present
$|C/T|$ is only bracketed between 0.08 and 0.37 \cite{MN}.  For comparison, the
corresponding $|C/T|$ ratio in $B^+ \to \bar D^0 \pi^+$ is about 0.4 \cite{NP}.
With the corresponding estimate $|C+T|/|T| = 1.23 \pm 0.15$, adding errors in
quadrature, we find $|t| \simeq |T| = (27.1 \pm 3.9) \times 10^{-9}$ GeV.  The
error associated with this estimate is superior to that obtained by applying
factorization to $B \to \pi l \nu$ \cite{Luo:2001ek}, which yields $|t| = (28.8
\pm 6.4) \times 10^{-9}$ GeV.  We shall use the former estimate for now.  An
improved estimate based on new $B \to \pi l \nu$ data \cite{CLEOsl} is in
progress \cite{LR03}.

A delicate point arises when passing from $T$ and $C$ to the $|\Delta S| = 1$
amplitudes $T'$ and $C'$.  In the combination $t'+c' = (T'+{P'}^C_{EW}) +(C' +
{P'}_{EW})$, the electroweak penguin terms contribute in magnitude about 2/3 of
the $|T'+C'|$ terms \cite{NR}.  Aside from an overall strong phase, one expects
\be
\label{eq:tplusc}
t'+c' = |T'+C'| \left[ e^{i \gamma} - \delta_{EW} \right],~~
\delta_{EW} = 0.65 \pm 0.15~~,
\ee
where the second term is the estimate of the electroweak penguin term.
$|T'+C'| = (9.35 \pm 0.70) \times 10^{-9}$ GeV is obtained by multiplying
$|t+c|\simeq |T+C|$ by the factor $|(\vus f_K)/(\vud f_\pi)| \simeq 0.280$,
with $\lambda = 0.2240$ \cite{Battaglia:2003in}.  The corresponding electroweak
penguin term contribution is $|T'+C'| \delta_{EW} = (6.1 \pm 1.5) \times
10^{-9}$ GeV.  It is expected to have the same weak phase as the
strangeness-changing penguin contribution $p'$ \cite{NR}.

We next extract the ``singlet penguin'' amplitude $|s'|$ by comparing the $B
\to \eta' K$ branching ratios with those expected on the basis of $p'$ alone.
The $p'$ contribution to $B \to \eta' K$ is much larger than that to $B \to
\eta K$ \cite{HJLP}, vanishing altogether for the latter for our choice $\eta =
(s \bar s - u \bar u - d \bar d)/\sqrt{3}$ as a result of cancellation of the
nonstrange and strange quark contributions.  In $\eta' = (2 s \bar s + u \bar u
+ d \bar d)/\sqrt{6}$ the nonstrange and strange quarks contribute
constructively in the $p'$ term, but not enough to account for the total
amplitude.  A flavor-singlet penguin term $s'$ added constructively to $p'$
with no relative strong phase and with $|s'/p'| \simeq 0.41$ can account for
the $B \to \eta' K$ decay rates.  We shall consider mainly a minimal $s'$ term
interfering constructively with $p'$, discussing in Sec.\ \ref{sec:amprel} the
possibility that $|s'|$ could be larger than its minimal value.  (The weak
phases of $p'$ and $s'$ are expected to be the same \cite{GR95,DGR}, but their
strong phases need not be.)

The amplitude for $B^+ \to \eta' K^+$ is better known than that for $B^0 \to
\eta' K^0$ (see Table \ref{tab:dS1}).  In the limit of $p',s'$ dominance they
should be equal, while that for the charged mode is slightly larger.  This
could be a consequence of a statistical fluctuation or a contribution from
$t'$.  The combination $t'+c'$ which appears in $A(B^+ \to \eta' K^+) = (92.4
\pm 2.7) \times 10^{-9}$ GeV includes an electroweak penguin term
$|T'+C'|\delta_{EW}/ \sqrt{6} = (2.5 \pm 0.6) \times 10^{-9}$ GeV which we
subtract from the total amplitude (the weak $p'$ and $s'$ phases are expected
to be $\pi$ as well) to obtain the estimate $|(3 p' + 4 s')/\sqrt{6}| = (89.9
\pm 2.8) \times 10^{-9}$ GeV.  In addition the term $|T'+C'|/\sqrt{6} = (3.8
\pm 0.3) \times 10^{-9}$ GeV contributes with unknown phase.  We thus combine
it in quadrature as an additional error to obtain $|(3 p' + 4 s')/\sqrt{6}| =
(89.9 \pm 4.7) \times 10^{-9}$ GeV from $B^+ \to \eta' K^+$.  We average this
value with $A(B^0 \to \eta' K^0) = (84.7 \pm 3.9) \times 10^{-9}$ GeV,
neglecting in the latter all $c'$ contributions including a possible
electroweak penguin term.  We then obtain $|(3 p' + 4 s')/\sqrt{6}| = (86.9 \pm
3.0) \times 10^{-9}$ GeV.  Assuming that $p'$ and $s'$ contribute
constructively as mentioned above, we subtract the $p'$ contribution to find
$s' = (18.9 \pm 2.2) \times 10^{-9}$ GeV.

The value of $|s/s'|$ is assumed to be governed by the same ratio of CKM
factors $|\vtd / \vts| = 0.197 \pm 0.012$ as $|p/p'|$, bearing in mind that the
full range of uncertainty including theoretical errors could be as much as
20\%.  We summarize the extracted magnitudes of amplitudes along with their
associated errors in Table \ref{tab:amp-value}.

Much theoretical effort has been expended on attempts to understand the
magnitude of the singlet penguin amplitude $s'$ \cite{Keta}.  An alternative
treatment \cite{BN} finds an enhanced standard-penguin contribution to $B \to
\eta' K$ without the need for a large singlet penguin contribution.  A key
feature of this work is the description of $\eta$--$\eta'$ mixing along the
lines of Ref.\ \cite{FKS}, involving a slightly different octet-singlet mixing
angle [$\theta_0 = (15.4 \pm 1.0)^\circ$ instead of our value of $19.5^\circ$].
The effect of the $s \bar s$ component of the wave function for both $\eta$ and
$\eta'$ is enhanced with respect to the symmetry limit.  We shall comment upon
one distinction between this scheme and ours at the end of the next section.
Predictions are given also for $|\Delta S| = 1$ decays involving one
pseudoscalar and one vector meson (see also Ref.\ \cite{Chiang:2001ir}), but
not for $\Delta S = 0$ decays.

\begin{table}
\caption{Values and errors of the topological amplitudes extracted according to
the method outlined in the text.
\label{tab:amp-value}}
\begin{ruledtabular}
\begin{tabular}{cc}
Amp. & Magnitude ($\times 10^{-9}$GeV) \\
\hline
$|t+c| \simeq |T+C| $  & $33.3 \pm 2.5$ \\
$|t| \simeq |T|$       & $27.1 \pm 3.9$ \\
$|c| \simeq |C|$       & $6.2 \pm 3.3$ \\
$|p|$                  & $9.00 \pm 0.64$ \\
$|s|$                  & $3.73 \pm 0.50$ \\
\hline
$|T'+C'|$              & $9.35 \pm 0.70$ \\
$|T'|$                 & $7.6 \pm 1.1$ \\
$|C'|$                 & $1.74 \pm 0.93$ \\
$|(T'+C')\delta_{EW}|$ & $6.1 \pm 1.5$ \\
$|p'|$                 & $45.7 \pm 1.7$ \\
$|s'|$                 & $18.9 \pm 2.2$ \\
\end{tabular}
\end{ruledtabular}
\end{table}

\section{$B \to \eta K$ AND $B \to \eta' K$ DECAYS \label{sec:Keta}}

The singlet penguin contribution to the $B \to \eta K$ amplitude is expected to
be $1/(2\sqrt{2})$ of that for $B \to \eta' K$, amounting to $(10.9 \pm 1.3)
\times 10^{-9}$ GeV.  As seen from Table \ref{tab:dS1}, this is an appreciable
fraction of the observed amplitude $A(B^+ \to \eta K^+) = (18.4 \pm 2.0) \times
10^{-9}$ GeV.  An additional electroweak penguin contribution of $|(T'+C')
\delta_{EW}|/\sqrt{3} = (3.50 \pm 0.85) \times 10^{-9}$ GeV leaves only $(4.0
\pm 2.5) \times 10^{-9}$ GeV to be accounted for via interference with
$|T'+C'|/\sqrt{3} = (5.4 \pm 0.4) \times 10^{-9}$ GeV.  This favors, but does
not prove, constructive interference between $|t'+c'|$ and $s'$.

Taking into account the $s'$ contribution alone (neglecting $c'$ including its
electroweak penguin part), one predicts $\ol{\cal B}(B^0 \to \eta K^0) = (1.03
\pm 0.24) \times 10^{-6}$.  We shall compare this with the current upper bound
in Sec.\ \ref{sec:unseen}.

We mentioned in Sec. \ref{sec:extract} that the value of $A(B^+ \to \eta'
K^+)$, after subtracting an electroweak penguin contribution, was $(89.9 \pm
2.8) \times 10^{-9}$ GeV, which is composed of the combination $(3 p' + 4
s')/\sqrt{6}$ [whose magnitude, averaging between charged and neutral modes, we
found to be $(86.9 \pm 3.0) \times 10^{-9}$ GeV], and a $T'+C'$ contribution
with magnitude $(3.8 \pm 0.3) \times 10^{-9}$ GeV.  Again, as in $B^+ \to \eta
K^+$, this favors but does not prove constructive interference between the
$|t'+c'|$ and penguin contributions.

Having now specified the necessary amplitudes, we can predict decay amplitudes
and $CP$ asymmetries for $B^+ \to \eta K^+$ and $B^+ \to \eta' K^+$, as well as
for the related process $B^+ \to \pi^0 K^+$, as functions of the CKM angle
$\gamma$ and a relative strong phase.  For the purpose of this discussion we
may write the decay amplitude for $B^+ \to M K^+$ ($M = \pi^0,~\eta,~\eta'$) as
\be
A(B^+ \to M K^+) = a(e^{i \gamma} - \delta_{EW})e^{i \delta_T} - b~~,
\ee
where the sign before $b$ takes account of the weak phase $\pi$ in the $|\Delta
S| = 1$ penguin term.  For $M = \pi^0,~\eta,~\eta'$ the values of $a$ are
$(6.61,~5.40,~3.82) \times 10^{-9}$ GeV, while those of $b$ are
$(32.32,~10.92,~86.85) \times 10^{-9}$ GeV, as one may see from the entries in
Table \ref{tab:dS1}.  The $CP$ rate asymmetries are
\be
{\cal A}_{CP}(f) \equiv \frac{|A(B^- \to \bar f)|^2 - |A(B^+ \to f)|^2}
{|A(B^- \to \bar f)|^2 + |A(B^+ \to f)|^2}~~,
\ee
while the $CP$-averaged amplitudes, to be compared with the experimental
amplitudes quoted in Tables \ref{tab:dS0} and \ref{tab:dS1}, are
\be
|A(f)| \equiv \left\{ \frac{1}{2} \left[ |A(B^+ \to f)|^2 + |A(B^- \to
 \bar f)|^2 \right] \right\}^{1/2}~~.
\ee
Here we have assumed the penguin and singlet penguin amplitudes $s'$ and $p'$
to have the same strong phase, which we take to be zero.

\begin{figure}
\includegraphics[height=4.2in]{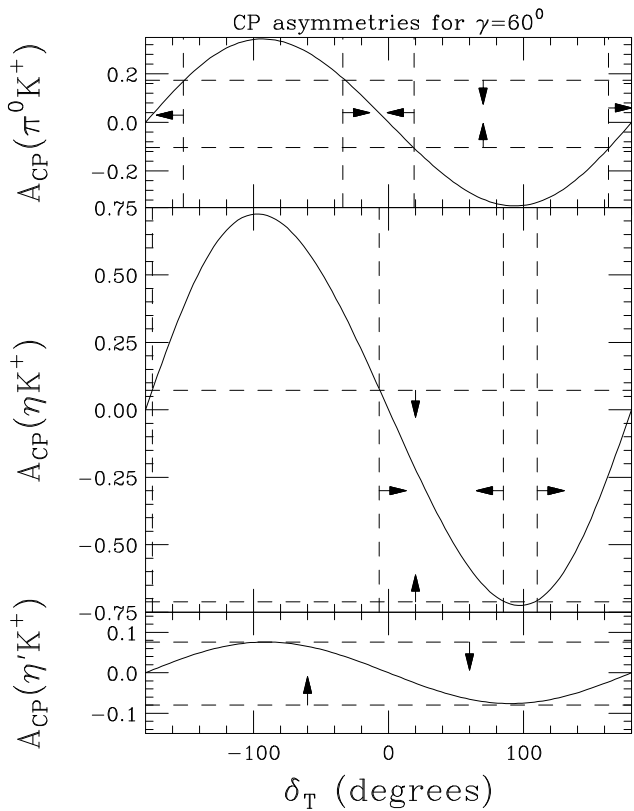}
\caption{Predicted $CP$ rate asymmetries when $\gamma = 60^\circ$ for $B^+ \to
  \pi^0 K^+$ (top), $B^+ \to \eta K^+$ (middle), and $B^+ \to \eta' K^+$
  (bottom).  Horizontal dashed lines denote 95\% c.l. ($\pm 1.96 \sigma$) upper
  and lower experimental bounds, leading to corresponding bounds on $\delta_T$
  denoted by vertical dashed lines.  Arrows point toward allowed regions.
\label{fig:acp_60}}
\end{figure}

The $CP$ asymmetries are most sensitive to $\delta_T$, varying less significantly
as a function of $\gamma$ over the 95\% c.l. allowed range \cite{CKMfitter}
$38^\circ < \gamma < 80^\circ$.  For illustration we present the asymmetries
calculated for $\gamma = 60^\circ$ in Fig.\ \ref{fig:acp_60}.

The constraints on $\delta_T$ from ${\cal A}_{CP}(\pi^0 K^+)$ are fairly
stringent: $-34^\circ \le \delta_T \le 19^\circ$ and a region of comparable
size around $\delta_T = \pi$.  The allowed range of ${\cal A}_{CP}(\eta K^+)$
restricts these regions further, leading to net allowed regions $-7^\circ \le
\delta_T \le 19^\circ$ or $163^\circ \le \delta_T \le 185^\circ$.  These
allowed regions do not change much if we vary $\gamma$ over its range between
$38^\circ$ and $80^\circ$.

\begin{figure}
\includegraphics[height=4.4in]{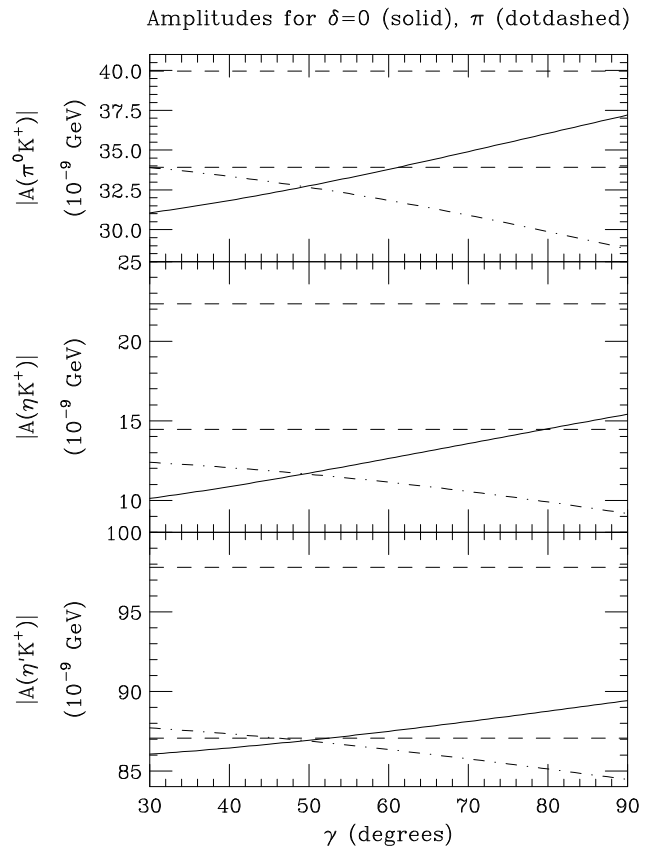}
\caption{Predicted magnitudes $|A|$ of amplitudes (based on $CP$-averaged rates)
  for $B^+ \to \pi^0 K^+$ (top), $B^+ \to \eta K^+$ (middle), and $B^+ \to
  \eta' K^+$ (bottom).  Solid and dot-dashed curves correspond to $\delta_T =
  0$ and $\pi$, respectively.  Horizontal dashed lines denote 95\% c.l. ($\pm
  1.96 \sigma$) upper and lower experimental bounds.
\label{fig:amps}}
\end{figure}

The predicted magnitudes $|A(f)|$ are very insensitive to $\delta_T$ within the
above ranges.  In Fig.\ \ref{fig:amps} we exhibit them for the two cases
$\delta_T = 0$ and $\delta_T = \pi$.  Values of $\delta_T$ near zero are
favored over those near $\pi$, and there is some preference for the higher
values of $\gamma$ within its standard model range.  The experimental value of
$|A(\eta K^+)|$ tends to exceed the prediction for all but the highest allowed
values of $\gamma$.

In contrast to our description of $B \to \eta K$ and $B \to \eta' K$, the
calculation of Ref.\ \cite{BN} has very small singlet penguin contributions to
both decays.  In the case of $B \to \eta' K$ a very small $CP$ asymmetry is
predicted as a result of the overwhelming dominance of the penguin amplitude.
The $B \to \eta K$ penguin amplitude does not vanish (in contrast to our
approach), but is predicted to be the dominant (small) contribution to the
decay, with a sign {\em opposite} to that in $B \to \eta' K$.  (See Table 3 of
Ref.\ \cite{BN}.)  Thus, for a given final-state phase, the $CP$ asymmetry
predicted in Ref.\ \cite{BN} for $B^+ \to \eta K^+$ will have the {\em opposite
  sign} to that which we predict.  This has interesting consequences for a
comparison with the $CP$ asymmetries in $B^+ \to \pi \eta$ and $B^+ \to \pi
\eta'$, which we will discuss in the next section.

A clear-cut difference between our formalism and that of Ref.\ \cite{BN} is in
the $CP$ asymmetries of $B \to \eta K$ and $B \to \eta' K$.  Assuming that the
singlet penguin amplitude has the same strong phase as the QCD penguin, we
predict both asymmetries to have the same sign for a fixed final-state phase of
penguin amplitudes relative to tree-level amplitudes.  However, the central
values of the predictions given in Ref.\ \cite{BN} favor the asymmetries to
have opposite signs.  Better measurements of ${\cal A}_{CP} (\eta K^+)$ and
${\cal A}_{CP} (\eta' K^+)$ (although the latter could be quite difficult) will
be very useful to justify which approach is more favored.

\section{CHARGED $\pi \eta^{(\prime)}$ MODES \label{sec:pi-eta}}

As seen in Table \ref{tab:dS0}, the magnitudes of $t+c$ and $p$ contributions
to the $\pi^{\pm} \eta^{(\prime)}$ modes are comparable to each other.  The CKM
factors associated with these amplitudes are $\vub^* \vud \propto e^{i \gamma}$
and dominantly $\vtb^* \vtd \propto e^{-i \beta}$, respectively.  One therefore
expects to observe sizeable direct $CP$ asymmetries in these decay modes if
there is a nontrivial relative strong phase in the amplitudes.  Indeed, a rate
asymmetry of $-0.51 \pm 0.19$ for $B^\pm \to \pi^\pm \eta$ has been observed at
BaBar \cite{Aubert:2003ez}.

On the other hand, the fact that the invariant amplitude prediction for $B^\pm
\to \pi^{\pm} \eta$ with both maximal constructive and destructive interference
schemes will be in conflict with the one extracted from experiments also
indicates a nontrivial phase between $t+c$ and $p$.
 
As outlined in Ref.~\cite{Chiang:2001ir}, combining the branching ratio and
$CP$ rate asymmetry information of the $\pi^{\pm} \eta$ modes, one should be
able to extract the values of the relative strong phase $\delta$ and the weak
phase $\alpha$, assuming maximal constructive interference between $p$ and $s$
(no relative strong phase).  The solution thus obtained can be used to predict
the branching ratio and $CP$ asymmetry of the $\pi^{\pm} \eta'$ modes.

Let us write the decay amplitudes for the $\pi^+ \eta$ and $\pi^+ \eta'$ modes
as
\bea
\label{eq:amppieta}
&&
A(\pi^+ \eta)
= -\frac{1}{\sqrt{3}}
  \left[ |t+c|e^{i\gamma} + |2p+s|e^{i(-\beta+\delta)} \right] ~, \\
&&
A(\pi^+ \eta')
= \frac{1}{\sqrt{6}}
  \left[ |t+c|e^{i\gamma} + |2p+4s|e^{i(-\beta+\delta)} \right] ~.
\eea
Then the $CP$ rate asymmetries ${\cal A}_{CP}(f)$
and the $CP$-averaged branching ratios
\begin{equation}
\overline{\cB}(f)
\equiv \frac{\cB(B^- \to \bar f) + \cB(B^+ \to f)}{2}
\end{equation}
are found to be
\bea
\label{eq:eq1}
{\cal A}_{CP}(\pi^+ \eta)
&\simeq& -\frac{0.91 \sin\delta \sin\alpha}{1 - 0.91 \cos\delta \cos\alpha} ~,\\
\label{eq:eq2}
{\cal A}_{CP}(\pi^+ \eta')
&\simeq& -\frac{\sin\delta \sin\alpha}{1 - \cos\delta \cos\alpha} ~, \\
\label{eq:eq3}
\overline{\cB}(\pi^+ \eta)
&\simeq& 4.95 \times 10^{-6}(1 - 0.91 \cos\delta \cos\alpha) ~, \\
\label{eq:eq4}
\overline{\cB}(\pi^+ \eta')
&\simeq& 3.35 \times 10^{-6}(1 - \cos\delta \cos\alpha) ~,
\eea
where the relation $\alpha = \pi - \beta - \gamma$ has been used and the
amplitudes have been substituted by the preferred values given in Table
\ref{tab:dS0}.

\begin{figure}[t]
\vspace{-1cm}
\includegraphics[width=3.8in]{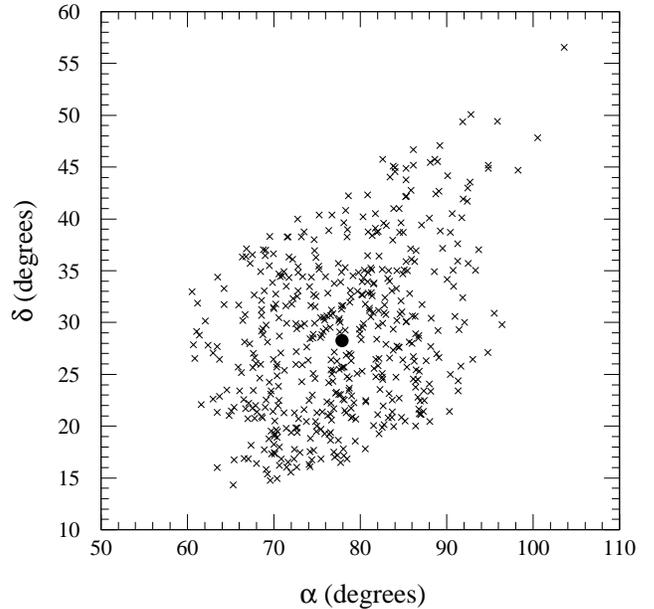}
\caption{Phases $\alpha$ and $\delta$ governing the decays $B^\pm \to
  \pi^\pm \eta$, obtained by solving constraints provided by branching ratio
  and direct $CP$ asymmetry
  along with amplitude inputs varying over allowed
  values, are depicted by scattered crosses.  The solution corresponding to
  the preferred central values is marked with a thick dot.
\label{fig:phase}}
\end{figure}

\begin{figure}[t]
\vspace{-1cm}
\includegraphics[height=3.8in]{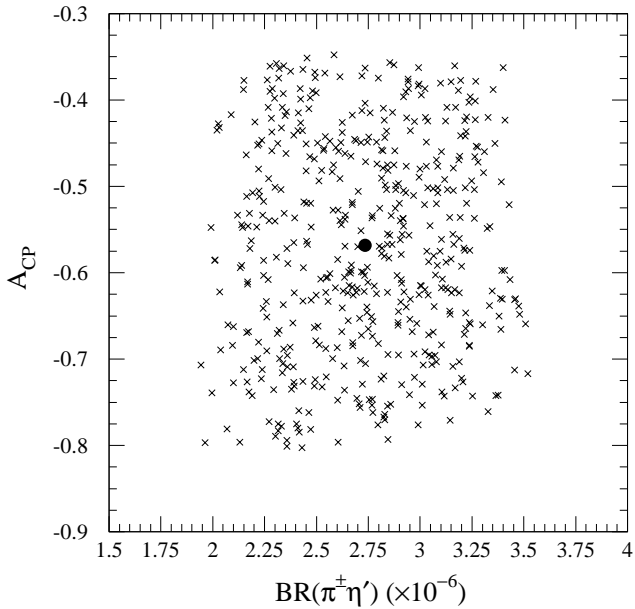}
\caption{Predicted values of the averaged branching ratio and direct $CP$
  asymmetry for the decays $B^\pm \to \pi^\pm \eta'$ corresponding to the
  points in Fig.\ \ref{fig:phase}.
\label{fig:value}}
\end{figure}

Note that Eqs.~(\ref{eq:eq1})--(\ref{eq:eq4}) are invariant under the exchange
$\alpha \leftrightarrow \delta$ and the transformation $\alpha \to \pi -
\alpha$ and $\delta \to \pi - \delta$.  (Although $\{ \alpha \to - \alpha,
\delta \to - \delta \}$ is also an invariant transformation, negative $\alpha$
is disfavored by current unitarity triangle constraints.)
In comparison, we have world averages of $\overline{\cB}(\pi^+ \eta) = (4.12
\pm 0.85)\times10^{-6}$ and ${\cal A}_ {CP}(\pi^+ \eta) = -0.51\pm0.19$.  We
use the central values to solve for the phases $\alpha$ and $\delta$ and obtain
four possibilities:
\be
(\alpha,\delta)
\simeq (78^{\circ},28^{\circ}) ~,
\ee
and those related by the $\alpha \leftrightarrow \delta$ and $(\alpha,\delta)
\to (\pi - \alpha,\pi - \delta)$ symmetries.  This information leads us to the
prediction of the branching ratio and $CP$ asymmetry for the $\pi^+ \eta'$
mode:
\bea
\overline{\cB}(\pi^+ \eta') &\simeq& 2.7 \times 10^{-6} ~, \\
{\cal A}_{CP}(\pi^+ \eta') &\simeq& -0.57 ~.
\eea

In general, we also allow the amplitude parameters ($|t+c|$, $|p|$, and $|s|$),
the branching ratio and direct $CP$ asymmetry of the $\pi^{\pm} \eta$ mode to
vary.  We use a normal distribution to sample 500 sets of input parameters
within the $1\sigma$ ranges as extracted in Sec.~\ref{sec:extract} and of the
experimental data.  For each set of input parameters, we go through similar
processes as outlined above to solve from ${\cal A}_{CP}(\pi^+ \eta)$ and
$\overline{\cB}(\pi^+ \eta)$ for the weak and strong phases.  They are found to
fall within the cross-marked area in Fig.~\ref{fig:phase}.  At $1\sigma$ level,
the weak phase $\alpha$ ranges from $\sim 60^{\circ}$ to $\sim 100^{\circ}$ and
the strong phase from $\sim 15^{\circ}$ to $\sim 55^{\circ}$.  As mentioned
before, there are three other possibilities related to Fig.~\ref{fig:phase} by
the $\alpha \leftrightarrow \delta$ and $(\alpha,\delta) \to (\pi - \alpha,\pi
- \delta)$ symmetries.  In either of these cases, the predicted values of
branching ratio and direct $CP$ asymmetry for the $\pi^{\pm} \eta'$ mode are
the same.  As shown in Fig.~\ref{fig:value}, the averaged branching ratio of
the $\pi^{\pm} \eta'$ modes is predicted to fall in the range $2.0 \times 10^
{-6}\alt \overline{\cB}(\pi^+ \eta') \alt 3.5 \times 10^{-6}$, which is well
below the best upper bound given in Table \ref{tab:dS0data}.  A sizeable direct
$CP$ asymmetry between $\sim -0.34$ and $\sim -0.80$ is expected from current
data.

\begin{figure}
\includegraphics[height=3.3in]{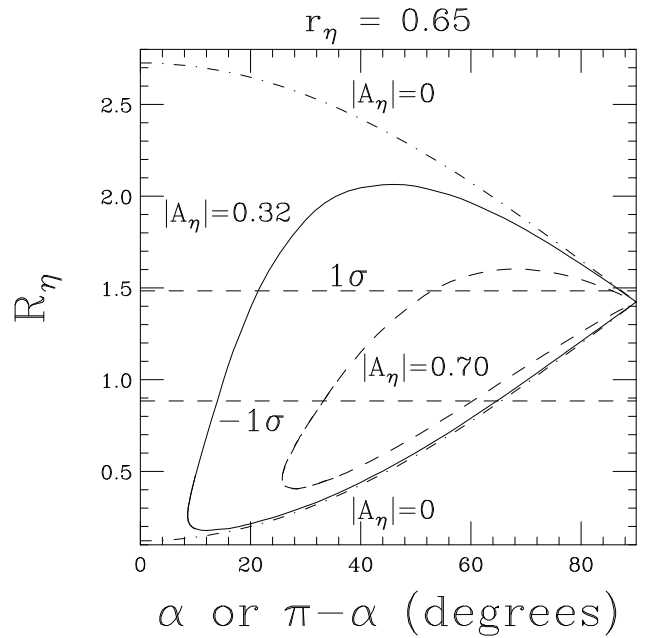}
\caption{Predicted value of $R_\eta$ (ratio of observed $CP$-averaged $B^\pm \to
\pi^\pm \eta$ decay rate to that predicted for tree amplitude alone) as a
function of $\alpha$ for various values of $CP$ asymmetry $|A_\eta|$.  (The
values 0.70 and 0.32 correspond to $\pm 1 \sigma$ errors on this asymmetry.)
\label{fig:Reta}}
\end{figure}

The amplitude relation (\ref{eq:amppieta}) and the corresponding
charge-conjugate amplitude may be written in the form
\bea
A(\pi^+ \eta) & = & -\frac{1}{\st}|t+c|e^{i \gamma} \left[ 1 - r_\eta e^{i
(\alpha + \delta)} \right]~~,\\
A(\pi^- \eta) & = & -\frac{1}{\st}|t+c|e^{-i \gamma} \left[ 1 - r_\eta e^{i
(-\alpha + \delta)} \right]~~,
\eea
where $r_\eta \equiv |2p+s|/|t+c| = 0.65 \pm 0.06$ is the ratio of penguin to
tree contributions to the $B^\pm \to \pi^\pm \eta$ decay amplitudes.  In
analogy with our previous treatments of $B^0 \to \pi^+ \pi^-$
\cite{GRpipi} and $B^0 \to \phi K_s$ \cite{phks}, we may define a quantity
$R_\eta$ which is the ratio of the observed $CP$-averaged $B^\pm \to \pi^\pm
\eta$ decay rate to that which would be expected in the limit of no penguin
contributions.  We find
\be
\label{eqn:Reta}
R_\eta = 1 + r_\eta^2 - 2 r_\eta \cos \alpha \cos \delta = 1.18 \pm 0.30~~.
\ee
One can then use the information on the observed $CP$ asymmetry in this mode to
eliminate $\delta$ and constrain $\alpha$.  (For a related treatment with a
different convention for penguin amplitudes see Ref. \cite{MGFPCP}.)  The
asymmetry is
\be
\label{eqn:Aeta}
A_\eta = -2 r_\eta \sin \alpha \sin \delta/R_\eta = -0.51 \pm 0.19~~,
\ee
so one can either use the simple result
\be
\label{eqn:RAeta}
R_\eta = 1 + r_\eta^2 \pm \sqrt{4 r_\eta^2 \cos^2 \alpha 
- (A_\eta R_\eta)^2 \cot^2 \alpha}
\ee
with experimental ranges of $R_\eta$ and $A_\eta$ or solve (\ref{eqn:RAeta})
for $R_\eta$ in terms of $\alpha$ and $A_\eta$.  The result of this latter
method is illustrated in Fig.\ \ref{fig:Reta}.

The range of $\alpha$ allowed at 95\% c.l.\ in standard-model fits to CKM
parameters is $78^\circ \le \alpha \le 122^\circ$ \cite{CKMfitter}.  For
comparison, Fig.\ \ref{fig:Reta} permits values of $\alpha$ in the three
ranges
\bea
14^\circ & \le & \alpha \le 53^\circ~~, \\
60^\circ & \le & \alpha \le 120^\circ~~, \\
127^\circ & \le & \alpha \le 166^\circ~~
\eea
if $R_\eta$ and $|A_\eta|$ are constrained to lie within their $1 \sigma$
limits.  These limits coincide with those extracted from Fig.\ \ref{fig:phase}
when one considers all the possible solutions related by symmetries.  Only the
middle range overlaps the standard-model parameters, restricting them very
slightly.  Better constraints on $\alpha$ in this region mainly would require
reduction of errors on $R_\eta$.

There are important questions of the consistency of the range of $\delta$ as
exibited in Fig.\ \ref{fig:phase} with other determinations of the relative
strong phases between penguin and tree amplitudes.  They involve comparisons
with two classes
of processes: (a) the $\Delta S = 0$ decays $B^0 \to \pi^+ \pi^-$, and (b)
the $|\Delta S|=1$ decays $B^+ \to (\pi^0,\eta,\eta')K^+$.

For $B^0 \to \pi^+ \pi^-$ the amplitude $t+p$ is not exactly the same as the
amplitude $A(B^+ \to \eta \pi^+) = -(t+c+2p+s)/\st$, which has small $c$ and
$s$ contributions and a larger penguin-to-tree ratio.  Nonetheless one should
expect the same sign of the $CP$ asymmetries ${\cal A}_{CP}(\eta \pi^+)$ and
${\cal A}_{\pi \pi}$, whereas the first is $-0.51 \pm 0.19$ while the second is
$0.51 \pm 0.19$.  It would be interesting to see whether explicit calculations
(e.g., using the methods of Refs.\ \cite{BBNS} and \cite{BN}) could cope with
this opposite sign.

In comparing the $|\Delta S|=1$ decays $B^+ \to (\pi^0,\eta,\eta')K^+$
discussed in Sec.\ \ref{sec:Keta} with $B^+ \to (\eta,\eta')\pi^+$ discussed in
the present
section, one expects in the flavor-$SU(3)$ limit that $\delta = - \delta_T$.
(We have associated the strong phase in each case with the less-dominant
amplitude: tree for $|\Delta S| = 1$ in Sec.\ \ref{sec:Keta} and penguin for
$\Delta S = 0$ in the present section.)  With the preference for $\delta > 0$
exhibited in Fig.\ \ref{fig:phase}, we would then expect to prefer $\delta_T
< 0$ in Fig.\ \ref{fig:acp_60}, which is disfavored by the negative central
value of the $CP$ asymmetry for $B^+ \to \eta K^+$.

To say it more succinctly, there is not a consistent pattern of direct $CP$
asymmetries within the present framework when one considers ${\cal A}_{CP}(\eta
\pi^+) < 0$ (favoring $\delta >0$) on the one hand, and both ${\cal A}_{\pi
  \pi}$ and ${\cal A}_{CP}(\eta K^+)$ (favoring $\delta < 0$) on the other
hand.  The measurement of a significant $CP$ asymmetry for $B^+ \to \eta'
\pi^+$ would provide valuable additional information in this respect.

As we mentioned at the end of the previous section, for a given final-state
phase we expect the calculation of Ref.\ \cite{BN} to give an opposite sign
to ours for ${\cal A}_{CP}(\eta K^+)$.  The current experimental central value
of ${\cal A}_{CP}(\eta K^+)$ (consistent with the range predicted in Ref.\
\cite{BN}) favors $\delta > 0$ in accord with ${\cal A}_{CP}(\eta \pi^+)$
which also favors $\delta > 0$ (Fig.\ 3).  It is then ${\cal A}_{\pi \pi}$
which is ``odd man out,'' favoring $\delta < 0$.

\section{$|\Delta S|=1$ CHARGED $B$ DECAYS AND THE RATIO $s'/p'$ 
\label{sec:amprel}}

Several relations among amplitudes were proposed in Refs.\ \cite{GR95} and
\cite{DGR} (see also \cite{DH}).  Notable among these was the quadrangle
relation for $B^+$ decay amplitudes
\be\label{quad}
A(\eta' K^+) = \sx A(\pi^+ K^0) + \st A(\pi^0 K^+) - 2 \s A(\eta K^+)~~.
\ee
We will show in the next section that this relation and a similar quadrangle
relation among $\Delta S=0$ amplitudes follows from U-spin symmetry alone.  A
quadrangle construction was suggested for $|\Delta S| =1$ processes and their
charge conjugates which permits the determination of the weak phase $\gamma$ as
long as the two quadrangles are not degenerate.  In order for this to be the
case, at least two of the three processes $B^+ \to \eta' K^+$, $B^+ \to \pi^0
K^+$, and $B^+ \to \eta K^+$ must have non-vanishing $CP$ asymmetries.  The
$CP$ asymmetry for $B^+ \to \pi^+ K^0$ must be very small if our assumption
that this decay is dominated by the penguin amplitude is correct.

We shall discuss a relation between $CP$-violating rate differences which
follows from the amplitude decompositions in Table \ref{tab:dS1}:
\bea
\label{eq:piplus}
A(\pi^+ K^0) & = & p'~~, \\
\s A(\pi^0 K^+) & = & -(p' + t' + c')~~,\\
\st A(\eta K^+) & = & -(s' + t' + c')~~,\\
\label{eq:etapr}
\sx A(\eta' K^+) & = & 3p' + 4s' + t' + c'~~.
\eea
We shall assume that the amplitudes $p'$ and $s'$ have the same weak phase but
not necessarily the same strong phase (in contrast to the simplified case
assumed in previous sections).  The amplitude $t'+c'$ has a weak phase $\gamma$
associated with its $T'+C'$ piece, and an electroweak penguin piece with the
same weak phase as $p'$ and $s'$.  Now let us define $CP$-violating rate
asymmetries $\Delta(f) \equiv \Gamma(\bar f) - \Gamma(f)$.  These may be
calculated by taking the difference between the absolute squares of the
amplitudes defined above and those for their charge-conjugate processes.  Under
the above assumptions about weak phases, we predict $\Delta(\pi^+ K^0) = 0$
(which is satisfied since ${\cal A}_{CP}(\pi^+ K^0) = -0.032 \pm 0.066$) and
\be
\Delta(\pi^0 K^+) + 2 \Delta(\eta K^+) = \Delta(\eta' K^+)~~.
\ee
This may be written in terms of observable quantities as
\bea
 && {\cal A}_{CP}(\pi^0 K^+) \ol\cB(\pi^0 K^+)
    + 2 {\cal A}_{CP}(\eta K^+) \ol\cB(\eta K^+) \nonumber \\
 && \qquad\qquad\qquad = {\cal A}_{CP}(\eta' K^+) \ol\cB(\eta' K^+)~~.
\eea
The individual terms in this equation (in units of $10^{-6}$) read
\be
(0.4 \pm 0.9) + (-2.0 \pm 1.3) = -0.2 \pm 3.1~~;
\ee
the sum on the left-hand side is $-1.6 \pm 1.6$.  The sum rule is satisfied,
but at least two terms in it must be individually non-vanishing to permit the
quadrangle construction of Ref.\ \cite{GR95}.  It does not make sense to
attempt such a construction with the present central values of the $CP$
asymmetries since they do not satisfy the sum rule exactly.

A related sum rule can be written for the rate asymmetries in $B \to \pi K$
decays.  Using similar methods, we find
\be
\Delta(\pi^0 K^0) = \frac{1}{2}\Delta(\pi^- K^+) - \Delta(\pi^0 K^+)~~.
\ee
This may be written as a prediction
\bea
{\cal A}_{CP}(\pi^0 K^0) 
&=& [\ol\cB(\pi^0 K^0)]^{-1}
    \left[ \frac{1}{2} {\cal A}_{CP}(\pi^- K^+) \ol\cB(\pi^- K^+) \right.
\nonumber \\
&&  \quad \left. - \frac{\tau_0}{\tau_+} {\cal A}_{CP}(\pi^0 K^+)
    \ol\cB(\pi^0 K^+) \right] \nonumber \\
&=& -0.11 \pm 0.08~~.
\eea

The most general check of our assumption that $p'$ and $s'$ have the same
strong phases (made in extracting the minimal value of $|s'|$ which would
reproduce the large $B \to \eta' K$ branching ratios) would rely on the
quadrangle construction of Ref.\ \cite{GR95}, which utilizes the rates for the
processes in Eqs.\ (\ref{eq:piplus})--(\ref{eq:etapr}) and their charge
conjugates.  As noted, in order to be able to perform this construction, one
must have quadrangles for processes and their charge conjugates which are of
different shapes, and thus (by virtue of the sum rule for rate differences) at
least two of the decays $B^+ \to \pi^0 K^+$, $B^+ \to \eta K^+$, and $B^+ \to
\eta' K^+$ must have non-zero $CP$ asymmetries.  Independently of whether such
asymmetries exist, one can still check the consistency of taking $s' = \mu p'$
(where $\mu$ is a real constant) by noting that under this assumption one has
$$
|A(\pi^+ K^0)|^2 (1+\mu)(1+2\mu)(1-\mu) + |A(\pi^0 K^+)|^2(1+\mu)
$$
\be
- |A(\eta K^+)|^2(1+2\mu) - |A(\eta' K^+)|^2(1-\mu) = 0~~.
\ee
Using the amplitudes quoted in Table \ref{tab:dS1}, one obtains the three
roots $\mu = (-2.21,~0.47,~1.24)$ to this cubic equation.  The value of
$|t'+c'|^2$ is a function of $\mu$ and squares of amplitudes.  Other ways
of writing $|t'+c'|^2$ give equivalent results.
$$
|t'+c'|^2 = |A(\pi^+ K^0)|^2(1 + 4 \mu + 2 \mu^2) + |A(\pi^0 K^+)|^2
$$
\be
+ 2 |A(\eta K^+)|^2 - |A(\eta' K^+)|^2~~.
\ee
The negative $\mu$ root gives negative $|t'+c'|^2$, while $\mu=1.24$ gives much
too large a value in comparison with the $|T'+C'|$ amplitude in Table
\ref{tab:amp-value}.  The root $\mu = 0.47 \pm 0.05$ is not far from the ratio
$s'/p' = 0.41 \pm 0.05$ implied by the values in Table \ref{tab:amp-value}.  It
implies a value of $|t'+c'| = (21.6 \pm 10.1) \times 10^{-9}$ GeV, still
somewhat large in comparison with the value $|T'+C'| = (9.53 \pm 0.70) \times
10^{-9}$ GeV but consistent with it given the large error and the uncertain
relative phase between $e^{i \gamma}$ and $\delta_{EW}$.

The error in the determination of $|t'+c'|$ using the above method is dominated
by that (22\%) in $\ol\cB(\eta K^0)$ (to be compared with 6--8\% in the other
three branching ratios).  To see the effect of a change in $\ol\cB(\eta K^+)$,
let us imagine that it is instead slightly below its present $1 \sigma$ limit,
or $(2.39 \pm 0.52) \times 10^{-6}$.  We then find $|t'+c'| = (9.4 \pm 19.2)
\times 10^{-9}$ GeV.  Alternatively, if $\ol\cB(\eta K^+)$ retains its present
central value but its error is decreased by a factor of 3 while the errors in
the other branching ratios remain the same, we find $|t'+c'| = (21.6 \pm 6.5)
\times 10^{-9}$ GeV.

If the error in $|t'+c'|$ as determined by the above method decreases to the
point that an inconsistency with Table \ref{tab:amp-value} develops, we would
be led to question at least one of the assumptions that (a) $\eta$ and $\eta'$
are the specific octet-singlet mixtures assumed here, and (b) the strong phases
of $s'$ and $p'$ are equal.

With improved knowledge of branching ratios and amplitudes one could extract a
relative strong phase between $s'$ and $p'$ from data.  In this approach,
instead of extracting $|t'+c'|$ from $|\Delta S| = 1$ $B^+$ decays as in the
above example, one would determine $|T'+C'|$ from $\Delta S=0$ transitions.
One also needs the relative size of the electroweak penguin $\delta_{EW}$, the
magnitude of $p'$ based on the $B^+ \to K^0 \pi^+$ decay rate, and the measured
$CP$-averaged branching ratios for $B^+ \to (\pi^0,\eta,\eta')K^+$.  With
these, one can solve for the magnitude and relative strong phase of $s'/p'$ and
the strong phase $\delta_T$ between the $T'+C'$ and $p'$ amplitudes.

Let $s' = \mu p' e^{i\delta_S}$ with $\mu > 0$ here and note that the weak
phase of the $p'$ amplitude is $\pi$ (as in Sec.\ \ref{sec:Keta}).  Then Eqs.\ 
(\ref{eq:piplus})--(\ref{eq:etapr}) may be rewritten as
\bea
A(\pi^+ K^0)
  & = & - |p'|~~, \\
\s A(\pi^0 K^+)
  & = & -\left[|T' + C'|(e^{i \gamma} - \delta_{EW}) e^{i
         \delta_T}  - |p'| \right]~~, \nonumber \\
&& \\
\st A(\eta K^+)
  & = & -\left[ |T' + C'|(e^{i \gamma} - \delta_{EW}) e^{i \delta_T}
         \right. \nonumber \\
&& \quad \left. - \mu |p'| e^{i\delta_S} \right]~~, \\
\sx A(\eta' K^+)
& = & |T' + C'|(e^{i \gamma} - \delta_{EW}) e^{i \delta_T} - 3|p'|
      \nonumber \\
&& \quad - 4 \mu |p'| e^{i\delta_S} ~~.
\eea
The first equation determines $|p'|$.  The remaining three then determine
$\mu$, $\delta_S$, and $\delta_T$ as functions of $\gamma$.

Taking the central values of the input parameters noted in the previous
paragraph, including $|p'| = 45.7 \times 10^{-9}$ GeV, $|T'+C'| = 9.7 \times
10^{-9}$ GeV and $\delta_{EW} = 0.65$, we find that $\gamma$ has to be greater
than $88^{\circ}$ in order to have solutions for $\mu$, $\delta_S$ and
$\delta_T$.  This feature arises from the need to reproduce the branching ratio
for $B^+ \to \pi^0 K^+$ which is slightly higher than expected on the basis of
penguin dominance.  One then needs maximal constructive interference between
the $|T'+C'|(e^{i \gamma} - \delta_{EW})$ and $-|p'|$ terms in the $B^+ \to
\pi^0 K^+$ amplitude, which forces $\gamma$ toward larger values.  This is the
basis of bounds originally presented in Ref.\ \cite{NR}.

To exhibit a less restrictive set of solutions, we take $\delta_{EW} = 0.80$
and the 95\% c.l. lower bound on the $B^+ \to \pi^0 K^+$ 
branching ratio, $\ol\cB \ge 10.7 \times 10^{-6}$.  The minimum value of
$\gamma$ permitting a solution is $51.9^\circ$. This is to be compared with the
result \cite{JRmor} based on consideration of all possible
errors on the ratio $2 \ol\cB(B^+ \to \pi^0 K^+)/ \ol\cB(B^+ \to \pi^+ K^0) =
1.30 \pm 0.15$: $\gamma \agt 58^\circ$ at the $1 \sigma$ level, and no lower
bound at 95\% c.l.

\begin{figure}
\includegraphics[width=3.5in]{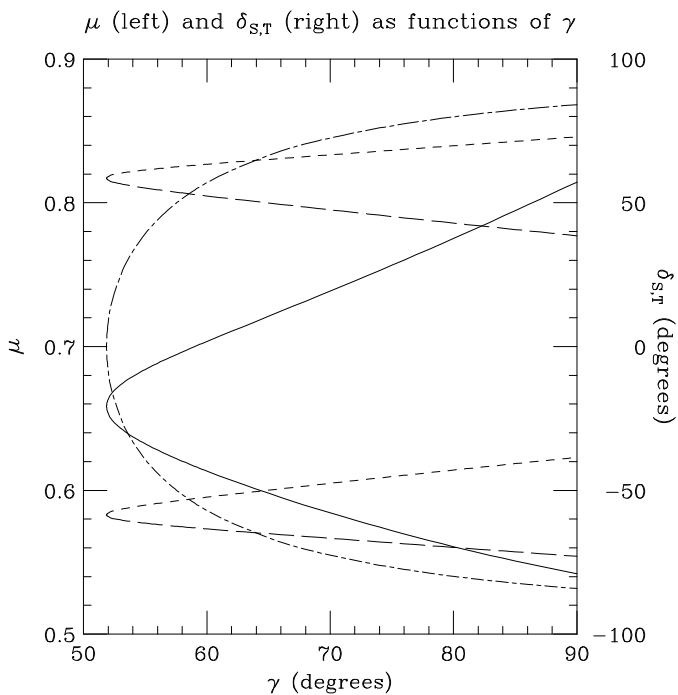}
\caption{Extracted values of $\mu \equiv |s'/p'|$ (solid) and strong phases
  $\delta_S$ (dashed) and $\delta_T$ (dash-dotted) as functions of the weak
  phase $\gamma$.  Positive (negative) values of $\delta_T$ are plotted with
  long (short) dash-dotted curves, with the corresponding $\delta_S$ values
  using long (short) dashed curves.  For either sign of $\delta_T$, the branch
  of $\delta_S$ with larger (smaller) absolute values corresponds to the upper
  (lower) branch of $\mu$.  Branches are joined at the point with minimum
  $\gamma$.  Here $\ol\cB(B^+ \to \pi^0 K^+) = 10.7 \times 10^{-6}$ and
  $\delta_{EW} = 0.80$.
\label{fig:mudelta}}
\end{figure}

\begin{figure}
\vspace{-0.3in}
\includegraphics[width=3.4in]{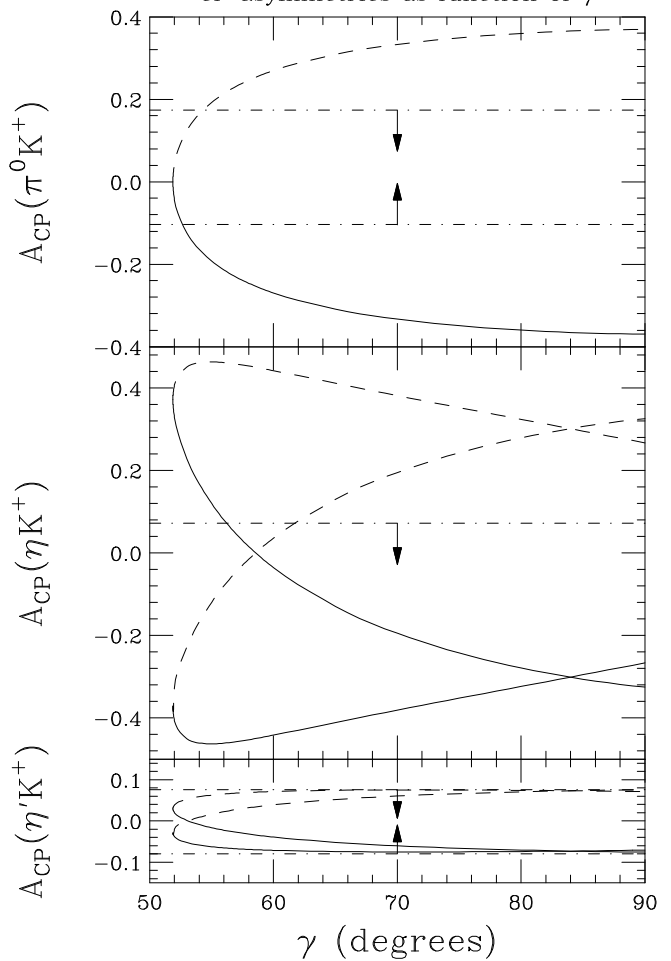}
\caption{Predicted $CP$ asymmetries for the $\pi^) K^+$, $\eta K^+$ and $\eta'
  K^+$ modes.  For all three plots, solid (dashed) curves correspond to the
  long dash-dotted positive (short dash-dotted negative) $\delta_T$ branch in
  Fig.\ \ref{fig:mudelta}.  The outer curves at low $\gamma$'s correspond to
  the branches of larger $|\delta_S|$ and larger $\mu$.  The corresponding
  $95\%$ c.l. bounds are also drawn in dash-dotted lines.  (The lower bound for
  ${\cal A}_{CP}(\eta K^+)$ is outside the plotting range.)  Here $\ol\cB(B^+
  \to \pi^0 K^+) = 10.7 \times 10^{-6}$ and $\delta_{EW} = 0.80$.
\label{fig:acp}}
\end{figure}

In Fig.\ \ref{fig:mudelta}, we show solutions for $\mu$, $\delta_S$ and
$\delta_T$ in the range $50^{\circ} \le \gamma \le 90^{\circ}$.  As $\gamma$
increases from its minimum, one obtains two branches of solutions for
$\delta_T$ differing by a sign.  Furthermore, there are two sets of possible
$\delta_S$ values for each sign of $\delta_T$, one set with larger absolute
values forming a branch that corresponds to larger values of $\mu$ while the
other with smaller absolute values forming the other branch that corresponds to
smaller values of $\mu$.  For a given $\mu$, $\delta_T \to - \delta_T$
corresponds to $\delta_S \to - \delta_S$.

With $\delta_{EW} = 0.80$, $\ol\cB \ge 10.7 \times 10^{-6}$, and the central
values of other input parameters, we find that the $CP$ asymmetry of the $\pi^0
K^+$ mode is predicted to be zero at the minimal value of $\gamma \simeq
51.9^{\circ}$, since the relative strong phase $\delta_T$ vanishes at that
point.  The $CP$ asymmetries of the $\eta K^+$ and $\eta' K^+$ modes at the
same value of $\gamma$, however, are predicted to be $\pm0.37$ and $\pm0.03$,
respectively.  The set of negative $CP$ asymmetries (corresponding to positive
$\delta_T$) is consistent with the current data as given in Table
\ref{tab:dS1}.  We plot the $CP$ asymmetries as functions of $\gamma$ in Fig.\ 
\ref{fig:acp}.  While the measured $CP$ asymmetry of the $\pi^0 K^+$ mode
\cite{Aubert:2003qj,Tomura} gives the strongest bound, $\gamma < 55^{\circ}$,
given the above-mentioned input conditions, this conclusion depends strongly on
the assumed branching ratios, particularly of the $\pi^0 K^+$ and $\eta K^+$
modes.  In any case it is clear that a solution is possible in principle for
both the relative magnitude and the relative phase of the singlet penguin and
ordinary penguin amplitude, given sufficiently reliable data.

\section{U-SPIN RELATIONS AMONG ALL CHARGED $B$ DECAYS \label{sec:uspin}}

While in previous sections we have employed the complete flavor $SU(3)$
symmetry group, neglecting small annihilation-type amplitudes, we will rely in
the present
section only on U-spin \cite{MLL,Uspin}, an important subgroup of $SU(3)$.
We will show that the eight charged $B$ decay amplitudes in Tables III and IV,
for both $|\Delta S|=1$ and $\Delta S=0$ transitions, are given in terms of two
triplets of U-spin amplitudes describing penguin and tree contributions.  This
implies several relations among these amplitudes including Eq.~(\ref{quad}) and
a similar quadrangle relation among $\Delta S=0$ amplitudes. Relations will
also be derived among penguin amplitudes in strangeness changing and
strangeness conserving decays, and among tree amplitudes in these decays. Such
relations may constrain tree amplitudes in $|\Delta S|=1$ decays and penguin
amplitudes in $\Delta S=0$ decays. Values calculated for these contributions in
previous sections, where stronger assumptions than U-spin were made, must obey
these constraints.

The U-spin subgroup of $SU(3)$ is the same as the I-spin (isospin) except that
the doublets with $U = 1/2, U_3 = \pm 1/2$ are
\be
\label{eqn:qks}
{\rm Quarks:}~~\left[ \begin{array}{c} |\half~~\half \rangle \\
                                      |\half -\! \half \rangle \end{array}
\right] = \left[ \begin{array}{c} |d \rangle \\ |s \rangle \end{array}
\right]~~,
\ee
\be
{\rm Antiquarks:}~~\left[ \begin{array}{c} |\half~~\half \rangle \\
                                      |\half -\!\half \rangle \end{array}
\right] = \left[ \begin{array}{c} |\bar s \rangle \\ -\!| \bar d \rangle
\end{array} \right]~~.
\ee
The charged $B$ is a U-spin singlet, while the charged kaon and pion belong 
to a U-spin doublet,
\be
|0~~0 \rangle = |B^+\rangle = |u\bar b\rangle~~,
\ee
\be
\left[ \begin{array}{c} |\half~~\half \rangle \\
                                      |\half -\!\half \rangle \end{array}
\right] = \left[ \begin{array}{c} |u\bar s \rangle = |K^+\rangle\\ 
-\!| u\bar d \rangle = -|\pi^+\rangle
\end{array} \right]~~.
\ee
Nonstrange neutral mesons belong either to a U-spin triplet or a U-spin
singlet.  The U-spin triplet residing in the pseudoscalar meson octet is
\be
\left[ \begin{array}{c} |1~~1 \rangle \\ |1~~0 \rangle \\ |1 -\!1 \rangle
\end{array} \right] = \left[ \begin{array}{c}
|K^0 \rangle = |d \bar s \rangle \\
\frac{\sqrt{3}}{2} |\eta_8 \rangle - \frac{1}{2} |\pi^0 \rangle
= \frac{1}{\sqrt{2}} | s \bar s - d \bar d \rangle \\
-\! |\ol K^0 \rangle = -\! |s \bar d \rangle \end{array} \right]~~,
\ee
and the corresponding singlet is
\be
|0~~0 \rangle = \frac{1}{2} |\eta_8 \rangle + \frac{\sqrt{3}}{2} |\pi^0 
\rangle = \frac{1}{\sqrt{6}} | s \bar s + d \bar d - 2u \bar u \rangle~~.
\ee 
In addition the $\eta_1$ is, of course, a U-spin singlet.  We take $\eta_8
\equiv (2 s \bar s - u \bar u - d \bar d)/\sqrt{6}$.  We shall also use $\eta_1
\equiv (u \bar u + d \bar d + s \bar s)/\sqrt{3}$, and recall our definition
$\pi^0 = (d \bar d - u \bar u)/\sqrt{2}$.

The $\Delta C=0,~\Delta S = 1$ effective Hamiltonian transforms like a $\bar s$
component ($\Delta U_3 = \half$) of a U-spin doublet, while the $\Delta C=0,
\Delta S = 0$ Hamiltonian transforms like a $-\bar d$ component ($\Delta U_3 =
-\half$) of another U-spin doublet.  Furthermore, one may decompose the two
Hamiltonians into members of the same two U-spin doublets multiplying given CKM
factors.  For practical purposes, it is convenient to use a convention in which
the CKM factors involve the $u$ and $c$ quarks, rather than the $u$ and $t$
quarks \cite{Uspin},
\bea
\label{Hs}
{\cal H}_{\rm eff}^{\bar b\to\bar s} & = & V^*_{ub}V_{us}O^u_s + 
V^*_{cb}V_{cs}O^c_s~~,\\
\label{Hd}
{\cal H}_{\rm eff}^{\bar b\to\bar d} & = & V^*_{ub}V_{ud}O^u_d + 
V^*_{cb}V_{cd}O^c_d~~.
\eea
Here $O_{d,s}^u$ and $O_{d,s}^c$ are two U-spin doublet operators, which for
simplicity of nomenclature will be called tree and penguin operators.

Since the initial $B^+$ meson is a U-spin singlet, the final states are U-spin
doublets, which can be formed in three different ways from the two U-spin
singlets and the U-spin triplet, each multiplying the U-spin doublet meson
states. Consequently, the eight decay processes can be expressed in terms of
three U-spin reduced matrix elements of the tree operator and three U-spin
penguin amplitudes.  Amplitudes corresponding to the U-spin singlet and triplet
in the octet and the $SU(3)$ singlet, will be denoted by $A^u_0,~A^u_1$ and
$B^u_0$, respectively for tree amplitudes and $A^c_0,~A^c_1$ and $B^c_0$ for
penguin amplitudes.  Complete $\Delta S=1$ and $\Delta S=0$ amplitudes for
U-spin final states made of two octets are given by
\bea
\label{eqn:As}
A^s_{0,1} & = & V^*_{ub}V_{us}A^u_{0,1} + V^*_{cb}V_{cs}A^c_{0,1}~~,\\
\label{eqn:Ad}
A^d_{0,1} & = & V^*_{ub}V_{ud}A^u_{0,1} + V^*_{cb}V_{cd}A^c_{0,1}~~.
\eea
Similar expressions for $B^s_0$ and $B^d_0$ describe decays to final states
involving $\eta_1$.

Absorbing a factor $1/2$ in the definition of $A^{u,c}_{0,1}$ one finds
\bea
\label{eqn:Uspin}
A(\eta_1K^+) & = & B^s_0~~,~~~~A(K^0\pi^+) = -\frac{4}{\sx}A^s_1~~,\\
A(\eta_8 K^+) & = & A^s_0 - A^s_1~~,\\
A(\pi^0 K^+) & = & \st A^s_0 + \frac{1}{\st} A^s_1~~,\\
A(\eta_1\pi^+) & = & B^d_0~~,~~~~A(\ol K^0 K^+) = -\frac{4}{\sx}A^d_1~~,\\
A(\eta_8\pi^+) & = & A^d_0 + A^d_1~~,\\
A(\pi^0\pi^+) & = & \st A^d_0 - \frac{1}{\st}A^d_1~~.
\eea

The physical $\eta$ and $\eta'$ states are mixtures of the octet and singlet.
In our convention if we write
\be
\label{eqn:etamix}
\eta = \cth \eta_8 - \sth \eta_1~~,~~~\eta' = \cth \eta_1 + \sth \eta_8~~,
\ee
the states defined in Sec.\ \ref{sec:not} correspond to $\cth \equiv \cos
\theta = 2 \sqrt{2}/3$, $\sth \equiv \sin \theta = 1/3$, $\theta = 19.5^\circ$.
Since the four physical $|\Delta S|=1$ amplitudes are expressed in terms of
three U-spin amplitudes, $B^s_0,~A^s_0$ and $A^s_1$, they obey one linear
relation given by Eq.~(\ref{quad}). Thus, this relation follows purely from
U-spin and does not require further approximations. A similar U-spin quadrangle
relation holds for $\Delta S =0$ amplitudes,
\be
\label{quad2}
A(\eta' \pi^+) = -\sx A(K^+ \ol K^0) + \st A(\pi^0 \pi^+) - 2 \s A(\eta 
\pi^+)~~.
\ee

Combining Eqs.~(\ref{eqn:As})--(\ref{eqn:Ad}) and
Eqs.~(\ref{eqn:Uspin})--(\ref{eqn:etamix}), one may relate penguin amplitudes
$A^c$ (or tree amplitudes $A^u$) in $\Delta S =0$ and $|\Delta S|=1$ decays. One
finds
\bea
\label{eqn:Aceta'}
A^c(\eta'\pi^+) & = &
A^c(\eta' K^+) -\frac{1}{\sx}A^c(K^0\pi^+)~~,\\
\label{eqn:Aceta}
A^c(\eta\pi^+) & = & 
A^c(\eta K^+)  - \frac{2}{\st}A^c(K^0\pi^+)~~,\\
\label{eqn:Aueta'}
A^u(\eta' K^+) & = & 
A^u(\eta'\pi^+) + \frac{1}{\sx}A^u(\ol K^0 K^+)~~,\\
\label{eqn:Aueta}
A^u(\eta K^+) & = & 
A^u(\eta \pi^+) + \frac{2}{\st}A^u(\ol K^0 K^+)~~.
\eea
We note that, because of the different conventions used here and in Tables III
and IV, the amplitudes $A^c$ and $A^u$ do not correspond {\it exactly} to
penguin and tree amplitudes in the Tables. It is straightforward to translate
amplitudes in one convention to the other convention \cite{conv}.

Let us focus our attention first on Eq.~(\ref{eqn:Aceta'}).  The two penguin
contributions on the right-hand-side dominate the corresponding measured
amplitudes. Therefore, the complex triangle relation implies a lower bound, at
90\% confidence level, on the penguin contribution to $B^+ \to \eta'\pi^+$ in
terms of measured amplitudes,
\bea
|V^*_{cb}V_{cd}A^c(\eta'\pi^+)|
&\ge&  \frac{|V_{cd}|}{|V_{cs}|} 
     \left[ |A(\eta' K^+)| - \frac{1}{\sx}|A(K^0\pi^+)| \right] \nonumber \\
&>& 16.1\times 10^{-9}~{\rm GeV}~~.
\eea
This should be compared with the tree contribution to this process,
\be
|V^*_{ub}V_{ud}A^u(\eta'\pi^+)| \approx \frac{1}{\st}|A(\pi^+\pi^0)|
= 13.6\times 10^{-9}~{\rm GeV}~~.
\ee
We conclude that {\it U-spin symmetry alone implies that the penguin
  contribution in $B^+\to\eta'\pi^+$ is at least comparable in magnitude to the
  tree amplitude of this process}. This confirms our more detailed estimate in
Section \ref{sec:pi-eta}. (The small differences between the lower bound and
this estimate follows from the different convention used and from the small
nonpenguin contribution in $B^+\to \eta'K^+$.)

Eq.~(\ref{eqn:Aceta}) implies a somewhat weaker lower bound on the penguin
contribution in $B^+\to\eta\pi^+$,
\bea
|V^*_{cb}V_{cd}A^c(\eta\pi^+)|
&\ge& \frac{|V_{cd}|}{|V_{cs}|}
    \left[- |A(\eta K^+)| + \frac{2}{\st}|A(K^0\pi^+)| \right] \nonumber \\
&>& 7.1\times 10^{-9}~{\rm GeV}~~.
\eea
Namely, the penguin amplitude in $B^+\to\eta\pi^+$ is at least $37\%$ of the
tree contribution to this process. This bound is weakened somewhat by using the
complete amplitude of $B^+\to \eta K^+$, which contains a sizable tree
amplitude.

Eqs.~(\ref{eqn:Aueta'}) and (\ref{eqn:Aueta}) may be used in order to obtain
upper bounds on tree contributions with weak phase $\gamma$ in $B^+ \to
\eta'K^+$ and $B^+ \to \eta'K^+$. Assuming that the physical amplitudes of
$B^+\to \eta\pi^+$ and $B^+\to \ol K^0 K^+$ are not smaller than the
corresponding tree amplitudes, one finds
\bea
&& |V^*_{ub}V_{us}A^u(\eta' K^+)| \nonumber \\
&& \qquad
\le \frac{f_K}{f_\pi} \frac{|V_{us}|}{|V_{ud}|}
    \left[ |A(\eta' \pi^+)| + \frac{1}{\sx}|A(\ol K^0 K^+)| \right]
\nonumber \\
&& \qquad < 9.1\times 10^{-9}~{\rm GeV}~~,\\
&& |V^*_{ub}V_{us}A^u(\eta K^+)| \nonumber \\
&& \qquad
\le \frac{f_K}{f_\pi} \frac{|V_{us}|}{|V_{ud}|}
    \left[ |A(\eta \pi^+)| + \frac{2}{\st}|A(\ol K^0 K^+)| \right] \nonumber \\
&& \qquad < 10.5\times 10^{-9}~{\rm GeV}~~.
\eea
These upper bounds imply that the tree contributions to $B^+\to \eta'K^+$ and
$B^+\to \eta K^+$ are less than 10\% and 66\% of the total amplitudes of these
processes, respectively. The bounds become 7\% and 42\% if one neglects the
small annihilation amplitude in $B^+\to \ol K^0 K^+$.

\section{MODES TO BE SEEN \label{sec:unseen}}

We summarize predicted branching ratios for some as-yet-unseen decay modes in
Table \ref{tab:predbrs}.

\begin{table}
\caption{Predicted branching ratios for some as-yet-unseen modes and present
90\% c.l.\ upper limits, in units of $10^{-6}$.
\label{tab:predbrs}}
\begin{ruledtabular}
\begin{tabular}{c l c c c}
  & Decay     & \multicolumn{2}{c}{Predicted} & Upper \\
  & mode      & This work & Ref.\ \cite{FHH} & limit \\ \hline
$B^+$ & $\to \pi^+ \eta'$ & $2.7 \pm 0.7$ \footnotemark[1] 
                          & $16.8^{+16.0}_{-9.7}$ & 7.0 \cite{Abe:2001pf} \\
  & $    \to K^+ \ol K^0$ & $0.75 \pm 0.11$ & $0.8^{+0.4}_{-0.2}$ 
                          & 1.3 \cite{Aubert:2002ng} \\
$B^0$ & $\to \pi^0 \pi^0$ & 0.4 to 1.6 \footnotemark[2]
                          & $1.9^{+0.8}_{-0.7}$ & 3.6 \cite{Aubert:2003qj} \\
  & $    \to \pi^0 \eta$  & $0.69 \pm 0.10$ & $1.2^{+0.6}_{-0.4}$ 
                          & 2.9 \cite{Richichi:1999kj} \\
  & $    \to \pi^0 \eta'$ & $0.77 \pm 0.11$ & $7.8^{+3.8}_{-4.3}$ 
                          & 5.7 \cite{Richichi:1999kj} \\
  & $    \to K^0 \ol K^0$ & $0.70 \pm 0.10$ & $0.7^{+0.4}_{-0.2}$ 
                          & 2.4 \cite{Aubert:2001hs} \\
  & $    \to \eta \eta$   & $0.3$ to $1.1$ \footnotemark[2]
                          & $3.1^{+1.3}_{-1.1}$ & 18 \cite{Hagiwara:fs} \\
  & $    \to \eta \eta'$  & $0.6$ to $1.7$ \footnotemark[2]
                          & $7.6^{+5.3}_{-3.4}$ & 27 \cite{Hagiwara:fs} \\
  & $    \to \eta' \eta'$ & $0.3$ to $0.6$ \footnotemark[2]
                          & $5.4^{+4.5}_{-3.1}$ & 47 \cite{Hagiwara:fs} \\
  & $    \to \eta K^0$    & $1.03 \pm 0.24$ & $2.4^{+0.5}_{-0.6}$ 
                          & 4.6 \cite{Aubert:2003ez} \\
\end{tabular}
\end{ruledtabular}
\footnotetext[1]{Predicted ${\cal A}_{CP} = -0.57 \pm 0.23$.}
\footnotetext[2]{Lower value from the central value of penguin amplitudes
  alone; upper value with constructive $c$--penguin interference and maximal
  $|c|$, $|p|$, and $|s|$ ($1\sigma$).}
\end{table}

We have already discussed the $B^+ \to \pi^+ \eta'$ mode in Sec.\ 
\ref{sec:pi-eta}.  The spread in the predicted charge-averaged branching ratio,
$\ol\cB(\pi^+ \eta') = (2.7 \pm 0.7) \times 10^{-6}$, reflects that shown in
Fig.\ \ref{fig:amps}.  The predicted $CP$ asymmetry is large: ${\cal
  A}_{CP}(\pi^+ \eta') =
-0.57 \pm 0.23$.  By contrast, Ref.\ \cite{FHH} finds $\ol\cB(\pi^+ \eta') =
(16.8^{+16.0} _{-9.7}) \times 10^{-6}$ (barely compatible with the upper limit
of $7 \times 10^{-6}$ in Table \ref{tab:dS0data}) and ${\cal A}_{CP}(\pi^+
\eta') = -0.18^{+0.15}_ {-0.09}$.

The prediction $\ol\cB(B^0 \to \eta K^0) = (1.03 \pm 0.24) \times 10^{-6}$
(discussed in Sec.\ \ref{sec:extract}) is given for the $s'$ contribution
alone.  It is a factor of about 2 below the value of $(2.4^{+0.5}_{-0.6})
\times 10^{-6}$ found in Ref.\ \cite{FHH}.

Using approximate $SU(3)_F$ symmetry, the amplitudes of both $B^+ \to K^+ \ol
K^0$ and $B^0 \to K^0 \ol K^0$ are the same ($p$) and related to the one
extracted from the $\pi^+ K^0$ mode.  Thus, their branching ratios are expected
to be $\sim 7.5 \times 10^{-7}$ and $\sim 7.0 \times 10^{-7}$, respectively.
These are rather close to central values (quoted with rather large errors) in
Ref.\ \cite{FHH}.  To observe these decay modes, the data sample should be
enlarged by a factor of $\sim 1.7$ and $\sim 3.4$.  These estimates do not
include additional possible theoretical errors on $p$ associated with the
methods of Sec.\ \ref{sec:extract}.

The decay $B^0 \to \pi^0 \pi^0$ receives contributions from the $p$ and $c$
amplitudes: $A(B^0 \to \pi^0 \pi^0) = (p-c)/\sqrt{2}$.  This amplitude is to be
compared with $A(B^0 \to \pi^+ \pi^-) = -(t+p)$, in which Table \ref{tab:dS0}
indicates that the tree and penguin amplitudes may be interfering
destructively.  Since one expects $c/t$ to be mainly real and positive
\cite{BBNS,MN}, one then expects either no interference or constructive
interference between $p$ and $c$ in $B^0 \to \pi^0 \pi^0$.  The $p$
contribution alone gives a branching ratio of about $0.4 \times 10^{-6}$, while
Table \ref{tab:amp-value} indicates that the $c$ contribution could be as large
as $p$.  If $c$ and $p$ then add constructively, one could have a branching
ratio as large as $1.6 \times 10^{-6}$.  This still lies below the present
upper limit, by a little more than a factor of 2.  A lower bound at the 90\%
c.l. of $\ol{\cal B}(B^0 \to \pi^0 \pi^0) \agt 0.2 \times 10^{-6}$ may be
obtained using the observed $B^+ \to \pi^+ \pi^0$ and $B^0 \to \pi^+ \pi^-$
branching ratios and isospin alone \cite{MGFPCP}.  For comparison, Ref.\ 
\cite{FHH}
predicts $\ol\cB(B^0 \to \pi^0 \pi^0) = (1.9^{+0.8}_{-0.7}) \times 10^{-6}$.
 
Since the $\pi^0 \eta^{(\prime)}$ modes involve linear combinations of $p$ and
$s$ that are believed to have the same weak phase and no sizeable relative
strong phase, we predict their branching ratios to be $(0.69 \pm 0.10) \times
10^{-6}$ and $(0.77 \pm 0.11) \times 10^{-6}$.  We therefore need about $4$ and
$7$ times more data than the sample on which the upper limits in Table VI are
based in order to see these decays.  This should not be difficult since those
limits were based on CLEO data alone \cite{Richichi:1999kj}.  The
corresponding predictions of Ref.\ \cite{FHH} are $\ol\cB(B^0 \to \pi^0 \eta) =
(1.2^{+0.6}_{-0.4}) \times 10^{-6}$ (slightly above ours) and $\ol\cB(B^0 \to
\pi^0 \eta') = (7.8^{+3.8}_{-4.3}) \times 10^{-6}$ (far above ours, with the
upper values excluded by experiment).

\section{SUMMARY \label{sec:summary}}

We have discussed implications of recent experimental data for $B$ decays into
two pseudoscalar mesons, with emphasis on those with $\eta$ and $\eta'$ in the
final states.  We present a preferred set of amplitude magnitudes in Tables
\ref{tab:dS0} and \ref{tab:dS1}, where quantities are either extracted directly
from data or related to one another by appropriate CKM and $SU(3)_F$ breaking
factors.  In particular, we make the assumption that the singlet penguin
amplitude and the QCD penguin in $|\Delta S| = 1$ transitions have the same
strong phase in the tables.  We show that this assumption is consistent with
current measurements of the branching ratios and $CP$ asymmetries of the
charged $B$ meson decays.  We also study the consequences of relaxing
this assumption but assuming that electroweak penguin contributions and
branching ratios are sufficiently well known.

We have extracted relative weak and strong phases between the tree-level
amplitudes and penguin-loop amplitudes in the $B^{\pm} \to \eta \pi^{\pm}$
modes, and shown how improved data will lead to stronger constraints.
Remarkably, branching ratio data can be at least as useful as $CP$ asymmetries
in this regard.  We use U-spin alone to argue for a large penguin contribution
in $B^+\to \eta'\pi^+$, and we predict a range of values for the branching
ratio and $CP$ asymmetry of this decay.  In particular, we predict $\ol\cB(B^+
\to \eta'\pi^+) = (2.7 \pm 0.7) \times 10^{-6}$ for the charge-averaged
branching ratio and, as a consequence of the apparent large $CP$ asymmetry in
$B^+ \to \eta\pi^+$, an even larger $CP$ asymmetry of ${\cal A}_{CP}(B^+ \to
\eta'\pi^+) = -0.57 \pm 0.23$.  We show that the present sign of the direct
$CP$ asymmetry in $B^+ \to \eta \pi^+$ conflicts with that in $B^0 \to \pi^+
\pi^-$ and, assuming flavor $SU(3)$, with that in $B^+ \to \eta K^+$.
[The $CP$ asymmetry
in $B^+ \to \eta K^+$ predicted by Ref.\ \cite{BN} would be opposite in sign
for the same final-state phase, agreeing with that in $B^+ \to \eta \pi^+$
and disagreeing with the direct asymmetry in $B^0 \to \pi^+ \pi^-$.]
Since none of these asymmetries has yet been established at the $3 \sigma$
level, there is not cause for immediate concern, but it would be interesting to
see whether any other explicit calculations (e.g., those of Ref.\
\cite{BBNS,BN}) for $B^+ \to \eta \pi^+$ can reproduce such a pattern.

Using $SU(3)$ flavor symmetry, we also have estimated the required data samples
to detect modes that have not yet been seen.  The one closest to being observed
is $B^+ \to K^+ \ol K^0$, which should be visible with about twice the
present number of observed $B$ decays.

\section*{ACKNOWLEDGMENTS}

We thank William Ford and James Smith for helpful discussions.
This work was supported in part by the United States Department of Energy, High
Energy Physics Division, through Grant Contract Nos.\ DE-FG02-90ER-40560 and
W-31-109-ENG-38.

\appendix

\section{NONPENGUIN CONTRIBUTIONS IN $B^0\to \eta' K_S$ \label{sec:B0eta'K}}

The angle $\beta$ can be measured through several different $B$ decay modes in
addition to the ``golden'' $B^0 \to J/\psi K_S$ channel and others involving
the $\bar b \to \bar c c \bar s$ subprocess.  The large branching ratio for
$B^0 \to \eta' K_S$ makes this mode particularly appealing; it is dominated in
our approach by the $\bar b \to \bar s$ penguin amplitude $p'$ and the
flavor-singlet penguin amplitude $s'$.  Within the standard model there are
several other possible contributions to this decay, including $c'$ in our
treatment and smaller amplitudes ($e',pa'$) which we neglect.

An estimate was performed \cite{GLNQ} with terms which could alter the
effective value of $\beta$ extracted from the $CP$-violating asymmetry
parameter ${\cal
  S}_{\eta' K_S}$.  While the full machinery of flavor $SU(3)$ was used, we
shall demonstrate that the U-spin subgroup employed in Section \ref{sec:uspin}
suffices.  We derive a linear relation among decay amplitudes differing from
that in Ref.\ \cite{GLNQ}, who neglected subtleties of symmetrization in
dealing with identical particles in an S-wave final state.  When these are
taken into account, the amplitudes listed in Tables \ref{tab:dS0} and
\ref{tab:dS1} satisfy the corrected linear relation.  Finally, we estimate the
corrections due to non-$(p',s')$ terms within our framework, finding them to be
much less important than in the more general treatment of Ref.\ \cite{GLNQ}.
In this respect we are much closer to the earlier approach of London and Soni
\cite{LS}, who concluded that such correction terms were insignificant.

Since we are considering S-wave decays of $B$ to two spinless final particle,
one must symmetrize the two-particle U-spin states.  The amplitudes in Tables
\ref{tab:dS0} and \ref{tab:dS1} are defined in such a way that their squares
times appropriate kinematic factors always give partial widths. For identical
particles, amplitudes satisfying Clebsch-Gordan relations are defined with
factors of $1/\s$ with respect to those in Tables \ref{tab:dS0} and
\ref{tab:dS1}.  We then reproduce results of Ref.\ \cite{GLNQ} using U-spin.
The first relation, written for our notation and phase convention, is
\be
\label{eqn:sing}
A^{u,c}(\eta_1 K^0) = \frac{1}{\sqrt{2}} A^{u,c}(\eta_1 \pi_0) - 
\sqrt{\frac{3}{2}}A^{u,c}(\eta_1 \eta_8)~~,
\ee
which refers to a single $U=1$ amplitude.  In the combination of $\pi^0$ and
$\eta_8$ on the right-hand side, the $U=0$ pieces cancel.  The $\eta_1$ is of
course a U-spin singlet.

The final state in $B^0 \to \eta_8 K^0$ involves $U = U_3 = 1$, since both
$|B^0 \rangle$ and the weak $|\Delta S| = 1$ Hamiltonian transform as
$|\half~~\half \rangle$.  The final states in $\Delta S = 0$ $B^0$ decays, on
the other hand, involve several possible U-spin combinations.  The total U-spin
can be either 0 or 1.  There are two ways of getting $U=0$: Each of the final
mesons can have either $U=0$ or $U=1$.  There is only one way of getting $U=1$:
One final meson must have $U=0$ and the other must have $U=1$.  This follows
from the symmetry of the final state; two $|1 ~~ 0 \rangle$ states cannot make
a
$|1 ~~ 0 \rangle$ state.  Thus there are three invariant amplitudes describing
four decays.  The appropriate relation between them is
\bea
\label{eqn:oct}
A^{u,c}(\eta_8 K^0) & = & \frac{1}{2}\sqrt{\frac{3}{2}}[A^{u,c}(\pi^0 \pi^0) 
- A^{u,c}(\eta_8\eta_8)] \nonumber\\
&& \quad - \frac{1}{2\sqrt{2}} A^{u,c}(\eta_8 \pi^0)~~.
\eea
Aside from some signs due to different conventions, this agrees with the result
of Ref. \cite{GLNQ}.

The physical $\eta$ and $\eta'$ states are given in Eq.~(\ref{eqn:etamix}) in
terms of the $\eta-\eta'$ mixing angle $\theta$.  Using this general
parametrization, and respecting the above mention symmetrization rule, we find
for $B^0 \to \eta' K^0$:
\bea\label{eqn:etap}
& & A^u(\eta' K^0) = \frac{2 \cth^2 - \sth^2}{2 \sqrt{2}} 
A^u(\eta' \pi^0) - \frac{3 \sth \cth}{2 \sqrt{2}} A^u(\eta \pi^0)\\
& &+ \frac{\sth}{2}\sqrt{ \frac{3}{2}}A^u(\pi^0 \pi^0) - \sqrt{\frac{3}{2}}
\left( 2 \sth \cth^2 + \frac{1}{2} \sth^3 \right) A^u(\eta' \eta')\nonumber\\
& & 
+ \frac{3}{2} \sqrt{\frac{3}{2}} \sth \cth^2 A^u(\eta \eta)
+ \sqrt{\frac{3}{2}}\left[ \cth \left( \frac{1}{2}\sth^2 - \cth^2 \right)
\right] A^u(\eta \eta')~~.\nonumber
\eea

Applying this relation in order to obtain an upper bound on the tree
contribution in $B^0\to \eta' K^0$ requires assuming that the amplitudes on the
right-hand-side dominate the corresponding processes. Using present upper
bounds on the magnitudes of these amplitudes would have led to a rather weak
bound, of about 40\% of the measured amplitude of $B^0\to \eta' K^0$. However,
Table III shows that this assumption cannot be justified. On the other hand,
estimating the $C'$ contribution to the amplitude using the Clebsch-Gordan
coefficients of Table \ref{tab:dS1} and the range quoted in Table
\ref{tab:amp-value}, we find it to be less than 1\%.  It is clear that
dynamical assumptions such as those made in Ref.\ \cite{LS} have considerable
effects in limiting non-penguin contributions to the decay $B^0 \to \eta' K^0$.

\end{document}